\documentclass[12pt, a4paper, twoside]{article}
 \usepackage[font=small,format=plain,labelfont=bf,up,textfont=normal,up,justification=justified,singlelinecheck=false]{caption}
 
\usepackage[a4paper, left=3cm,right=2cm]{geometry}
\usepackage{hyperref}
\usepackage{amsfonts} 
\usepackage{amsmath}  
\usepackage{setspace}   
\usepackage{color}  
 
\usepackage[utf8,applemac]{inputenc} 
\usepackage{tensor}
\usepackage{cite}
\usepackage{tikz}
\usepackage{graphicx}
\usepackage{dcolumn}
\usepackage{bm}

\newcommand{\bea}{\begin{eqnarray}}
\newcommand{\eea}{\end{eqnarray}}
\newcommand{\be}{\begin{equation}}
\newcommand{\ee}{\end{equation}}
\def\nn{\nonumber}

\def\p{\partial}
\def\eps{\epsilon}

  \newcommand{\beqs}{\begin{eqnarray}}
\newcommand{\eeqs}{\end{eqnarray}}

\title{Vacua of the gravitational field}

\begin{document}


\setcounter{tocdepth}{1}

\begin{titlepage}

\begin{flushright}\vspace{-3cm}
{\small
\today }\end{flushright}
\vspace{0.5cm}

\begin{center}
{ \LARGE{\bf{Classical static final state of collapse\\  \vspace{2mm} with supertranslation memory}}}
 \vspace{5mm}

\vspace{2mm} 
\centerline{\large{\bf{Geoffrey Comp\`{e}re\footnote{e-mail: gcompere@ulb.ac.be}, Jiang Long\footnote{e-mail:
Jiang.Long@ulb.ac.be}}}}

\vspace{2mm}
\normalsize
\bigskip\medskip
\textit{Universit\'{e} Libre de Bruxelles and International Solvay Institutes\\
CP 231, B-1050 Brussels, Belgium
}

\vspace{25mm}

\begin{abstract}
\noindent
{The Kerr metric models the final classical black hole state after gravitational collapse of matter and radiation. Any stationary metric which is close to the Kerr metric has been proven to be diffeomorphic to it. 
Now, finite supertranslation diffeomorphisms are symmetries which map solutions to inequivalent solutions  as such diffeomorphisms generate conserved superrotation charges. The final state of gravitational collapse is therefore parameterized by its mass, angular momentum and supertranslation field, signaled by its conserved superrotation charges.

In this paper, we first derive the angle-dependent energy conservation law relating the asymptotic value of the supertranslation field of the final state to the details of the collapse and subsequent evolution of the system. We then generate the static solution with an asymptotic supertranslation field and we study some of its properties. Up to a caveat, the deviation from the Schwarzschild metric could therefore be predicted on a case-by-case basis from accurate modeling of the angular dependence of the ingoing and outgoing energy fluxes leading to the final state. 
}

\end{abstract}


\end{center}

\end{titlepage}

\newpage
\tableofcontents

\section{Introduction}

Strong astronomical evidence points to the existence of a supermassive object at the center of each galaxy including our own Milky Way whose radius is comparable to the event horizon of a black hole \cite{2008Natur.455...78D}. Black holes are the only known classical solutions to Einstein's equations which can form as a result of gravitational collapse of matter and radiation. Indeed, no state of matter or radiation has been found which could withstand the pressure of gravitational pull beyond a certain mass density \cite{1939PhRv...55..364T,1939PhRv...55..374O}, as also confirmed in recent numerical simulations \cite{Pretorius:2004jg,Choptuik:2015mma}. These simulations strengthen the cosmic censorship conjecture  \cite{Penrose:1969pc,Hawking:1976ra} which states that naked singularities never form in Einstein gravity coupled to matter, except in extremely fine-tuned and therefore unphysical scenarios. 

Black holes are however plagued with the black hole information paradox \cite{Hawking:1974sw}. The paradox arises from the following statements which are all strongly supported but which seem inconsistent: 1. Stationary black holes describe the final state of gravitational collapse; 2. The classical metric of stationary black holes is isometric to the Kerr black hole which is uniquely fixed by the mass and angular momentum   \cite{Hawking:1973uf,Carter:1971zc,Robinson:1975bv,Chrusciel:2008js,Alexakis:2009ch}; 3. The Kerr black hole emits an exactly thermal radiation without detailed quantum information which leads the black hole to evaporate \cite{Hawking:1974sw}; 4. Remnants, i.e. massless objects with an arbitrary large entropy, do not exist (for a review see \cite{Chen:2014jwq}); 5. Information cannot be destroyed (or in other words, physical processes are unitarity). No definite resolution of this paradox has been proposed so far. Arguments based on quantum information theory show that a large effect has to exist close to the would be event horizon of the black hole for allowing information about the collapse to be emitted \cite{Mathur:2009hf} (see also \cite{Braunstein:2009my,Almheiri:2012rt,Bena:2013dka}). 

A possible resolution of the black hole information paradox was suggested in \cite{Strominger:2014pwa,Pasterski:2015tva,Hawking:2016msc} using the fundamental symmetry structure of asymptotically flat gravity: BMS symmetries\footnote{The details of our reasoning differ from \cite{Hawking:2016msc}. We will {argue} that black holes classically finitely deviate from the Schwarzschild metric (at order $\hbar^0$ not $\hbar^1$) mainly because memory effects are classical and finite. Non-linearities will be taken into account in the bulk of the spacetime. Classical supertranslation charges are zero but classical superrotation charges are non-trivial.}. These symmetries were originally found as asymptotic symmetries of gravity at future null infinity by Bondi, van der Burg, Metzner and Sachs \cite{Bondi:1962px,Sachs:1962wk}. The BMS asymptotic symmetries form an algebra isomorphic to a semi-direct sum of the Lorentz algebra with an abelian normal subgroup: the supertranslations which generalize the translations. In 2009, two new features of BMS symmetries were pointed out by Barnich and Troessaert \cite{Barnich:2009se} inspired from earlier work \cite{Sachs:1962zza,Penrose:1962ij}: the BMS algebra can be embedded in a natural \emph{enhanced} BMS algebra which also contain superrotations, singular supertranslations and Weyl rescalings. Also, the asymptotic BMS algebra can be extended in the bulk of the spacetime as a formal infinite series expansion using a generalized Lie bracket between the symmetry generators. The enhanced BMS algebra and its realization by associated canonical charges in the solution space of Einstein gravity was further investigated in \cite{Barnich:2011ct,Barnich:2010eb,Barnich:2011mi,Barnich:2013axa,Barnich:2016lyg}.

New connections between the original and enhanced BMS algebra and the physics of gravity and matter in asymptotically flat spacetimes were recently pointed out. The S-matrix is invariant under the original BMS symmetry which should act simultaneously on the future and past null infinity \cite{Strominger:2013jfa}.  In the case of small non-linear perturbations of Minkowski spacetime, the fields can be related at the past of the future null boundary with the fields at the future of the past null boundary via the antipodal map of points on the boundary sphere \cite{Strominger:2013jfa} after assuming suitable boundary conditions \cite{Christodoulou:1993uv}. In the semi-classical approximation, quantized gravitons amplitudes obey Weinberg's soft graviton theorem \cite{Weinberg:1965nx} and the subleading soft graviton theorem \cite{Cachazo:2014fwa} which can be understood as Ward identities of (enhanced) BMS symmetries \cite{He:2014laa,Kapec:2014opa,Campiglia:2014yka}. 

Most importantly for our concerns, the classical gravitational memory effect \cite{Zeldovich:1974aa,Braginsky:1986ia,Braginsky:1987aa,Ludvigsen:1989kg,Christodoulou:1991cr} was understood to be a consequence of the net BMS supertranslation shift arising from the total matter and radiative energy flux passing through the detector at future null infinity \cite{Geroch:1981ut}. The relationship between BMS symmetry and memory effects was made more precise in \cite{Strominger:2014pwa,Pasterski:2015tva}. 

The physical picture that emerges is that the metric contains \emph{three distinct but coupled fields}. It contains the Newtonian field, corrected by relativistic effects, which is responsible for gravitational attraction. It contains the gravitational wave field which can be decoupled in linearized gravity around Minkowski spacetime and which consists of two local propagating degrees of freedom of spin 2. The gravitational wave field is coupled to the Newtonian field at the non-linear level. The metric however also contains another field which we will call the \emph{supertranslation field} which can be best isolated at null infinity. The supertranslation field transforms under BMS supertranslation symmetries which map a physical state to another physical state\footnote{One cannot define a BMS superrotation field since it would otherwise lead to a vacuum with an unbounded energy spectrum \cite{Compere:2016jwb}.}. In this picture, the interactions between the supertranslation field and matter or gravitational radiation at future null infinity are responsible for the memory effects. Uniqueness theorems state that stationary analytic spacetimes are diffeomorphic to the Kerr metric \cite{Hawking:1973uf,Carter:1971zc,Robinson:1975bv,Chrusciel:2008js}.\footnote{The hypothesis of analyticity can be traded for closeness to the Kerr metric \cite{Alexakis:2009ch}.} But diffeomorphisms contain physical supertranslations. Therefore, a general stationary metric contains both the Newtonian field and the supertranslation field. The final state of gravitational collapse is therefore determined by mass, angular momentum and the final supertranslation field.

The nature of the supertranslation field in the bulk spacetime away from null infinity has only been recently investigated \cite{Compere:2016jwb}. Instead, three dimensional Einstein gravity as a toy model has been much more studied and understood. The lower dimensional analogue of the supertranslation field can be defined in the bulk spacetime \cite{Barnich:2010eb}. It transforms under the asymptotic symmetry algebra which consists of (both regular)   supertranslations and superrotations \cite{Ashtekar:1996cd,Barnich:2006av}. In the presence of a negative cosmological constant, the analogous field is the holographic stress-tensor which consists of a left and a right moving function on the unit cylinder \cite{Banados:1998gg} and which transforms under two copies of the Virasoro algebra \cite{Brown:1986nw}. The precise embedding of the holographic stress-tensor in the metric depends upon the choice of gauge. This embedding has been understood in Fefferman-Graham gauge \cite{Henningson:1998gx} and in the null Gaussian gauge \cite{Barnich:2012aw}. Infinitesimal conformal transformations of the stress-tensor are understood in the AdS/CFT correspondence \cite{Maldacena:1997re} as a physical change of state in the dual CFT.  In the large AdS radius limit, some components of the holographic stress-tensor admit a suitable limit which is the supertranslation field \cite{Barnich:2012aw}. The two copies of the Virasoro algebra then reduce to the BMS algebra including its central extension \cite{Barnich:2006av}. In three dimensional Einstein gravity without cosmological constant there is no black hole \cite{Ida:2000jh}. There is however a black hole in $AdS_3$, the BTZ black hole \cite{Banados:1992wn,Banados:1992gq}. The BTZ black hole can be equipped with a holographic stress-tensor. This leads to a black hole whose horizon is finitely deformed by the left moving and right moving components of the holographic stress-tensor \cite{Banados:1998gg}. Finally this deviation can be measured by conserved charges defined in the vicinity of the black hole because the charges are conserved under any smooth deformation of the circle from infinity up to the horizon \cite{Compere:2014cna,Compere:2015knw}. It can therefore be expected from these three dimensional models that four dimensional black holes equipped with a classical supertranslation field will admit a finite classical departure from the Kerr metric in the bulk, signaled by canonical superrotation charges.

Two main questions arise. First, one needs to determine the value of the supertranslation field in the final state of gravitational collapse due to an arbitrary incoming matter and radiation flux and after taking into account the total outgoing radiation flux. We will obtain the final value of the asymptotic supertranslation field  up to a caveat following closely the reasoning of \cite{Strominger:2014pwa,Pasterski:2015tva}. The second question is how to generate the Kerr solution equipped with an arbitrary supertranslation field. This task requires a finite BMS supertranslation diffeomorphism which we will present in this paper. For technical simplicity, we will restrict ourselves to the static case and only closely study the Schwarzschild metric equipped with a supertranslation field generated by a finite BMS supertranslation diffeomorphism.

\section{Supertranslation field in the final state}
\label{asf}

An angular-dependent conservation of energy equation between past and future null infinity was derived in \cite{Strominger:2014pwa,Pasterski:2015tva} for Christodoulou-Klainerman spacetimes \cite{Christodoulou:1993uv}. Here, we define boundary conditions which allow to extend this conservation law in the case of gravitational collapse leading to a massive stationary state.  It allows to relate the amplitude of the asymptotic supertranslation field in the final state as compared with the initial state with the leading energy profile of the collapsing matter and radiation at null infinity and a boundary term at spatial infinity. We will then draw the consequences of this conservation law in the case of the collapse of ideal spherical and non-spherical null shells under the assumption that the boundary term at spatial infinity vanishes, which is our main caveat. 

We consider a general asymptotically flat spacetime at future and past null infinity in the original sense of BMS \cite{Bondi:1962px,Sachs:1962wk}. We denote the coordinates around $\mathfrak I^+$ as $(u,r,z,\bar z)$ and around $\mathfrak I^-$ as $(v,r,w,\bar w)$. We denote the asymptotic past of $\mathfrak I^+$ as $\mathfrak I^+_-$ and its asymptotic future as $\mathfrak I^+_+$. We define similarly $\mathfrak I^-_\pm$. The expansions read as 
\bea
ds^2 &=& -du^2-2du dr + 2r^2 \gamma_{z \bar z} dz d\bar z + \frac{2m_B}{r}du^2 + r C_{zz}dz^2 + \dots \label{BC}\\
  &=& -dv^2 + 2dv dr + 2r^2 \gamma_{w \bar w} dw d\bar w + \frac{2m_B}{r}dv^2 + r C_{ww}dw^2 + \dots
\eea
Here $(z,\bar z)$ are stereographic coordinates ($z = e^{i \phi}\cot\frac{\theta}{2}$, $\gamma_{z \bar z}=\frac{2}{(1+z \bar z)^2}$). We identify points on $\mathfrak I^+_-$ and $\mathfrak I^-_+$ via the antipodal map $w = -\bar z^{-1}$, $\bar w = -z^{-1}$. 


At future null infinity, the Bondi mass aspect $m_B(u,z,\bar z)$ and the field $C_{zz}(u,z,\bar z)$ (whose $u$-independent mode depends upon the supertranslation field $C(z,\bar z)$ as we will shortly see) are related through Einstein's equations $G_{\mu\nu} = 8 \pi G T^M_{\mu\nu}$ to the total energy flux at null infinity $T_{uu}(u,z,\bar z)$ as
\bea
\p_u \left( m_B - \frac{1}{4} (D_z^2 C^{zz} + D_{\bar z}^2 C^{\bar z \bar z} )\right) = - T_{uu},\label{c1}\\
T_{uu} \equiv \frac{1}{4}N_{zz}N^{zz} + 4 \pi G \lim_{r \rightarrow \infty} [r^2 T_{uu}^M].
\eea
The Bondi mass decreases with retarded time due to gravitational radiation and null matter leaving the bulk through null infinity. 
At past null infinity, the same reasoning leads to
\bea
\p_v \left( m_B - \frac{1}{4} (D_w^2 C^{ww} + D_{\bar w}^2 C^{\bar w \bar w} )\right) = + T_{vv},\label{c2}\\
T_{vv} \equiv \frac{1}{4}N_{ww}N^{ww} + 4 \pi G \lim_{r \rightarrow \infty} [r^2 T_{vv}^M].
\eea
The Bondi mass increases with advanced time due to gravitational radiation and null matter entering the bulk through null infinity.

Let us now consider the gravitational collapse of a massive body. We may include both incoming gravitational radiation and null matter flux from $\mathfrak I^-$ and initial matter at past timelike infinity $i_-$. Gravitational radiation and null matter escapes at $\mathfrak I^+$. We assume that the spacetime reaches a stationary final state asymptotically at large $ |u|$ and $ |v|$. Precise such boundary conditions were formulated in \cite{Christodoulou:1993uv} for spacetimes with small (non-linear) deviations from Minkowski spacetime. Such boundary conditions lead to a final state at $\mathfrak I^+_+$ with vanishing Bondi mass. Since we want to consider a massive final state, we need more general boundary conditions. We assume the same fall-off conditions on radiative fields as \cite{Christodoulou:1993uv}. In particular, we take the news tensor $N_{zz}=\p_u C_{zz}$ to obey  $N_{zz} \sim |u|^{-3/2}$ and $N_{ww} \sim  |v|^{-3/2} $ for $|u|,|v| \rightarrow \infty$. 

Let us integrate \eqref{c1} between $-\infty$ and $+\infty$. We denote the final stationary state mass by $M_{stat} = \lim_{u \rightarrow \infty}m_B(u)$. We will show in a moment that it does not depend upon $z,\bar z$. Similarly, the Bondi mass at $u = - \infty$ is the total mass of the system $M_{tot}$.  Using \eqref{c1}, the energy radiated away is therefore
\bea
M_{tot} -M_{stat} =- \frac{1}{4} \left[ D_z^2 C^{zz} + D_{\bar z}^2 C^{\bar z \bar z} \right]^{\infty}_{u=-\infty} +  \int_{-\infty}^{\infty} du T_{uu}  . \label{c3}
\eea
Now, for center-of-mass Christodoulou-Klainerman spacetimes \cite{Christodoulou:1993uv}, one has the boundary conditions
\bea
D_z^2 C^{zz} - D_{\bar z}^2 C^{\bar z \bar z} |_{\mathfrak I^+_\pm}=0.\label{BC1}
\eea
It is implied by  the vanishing of the imaginary part of the Weyl tensor component $\Psi^0_2$ at $\mathfrak I^+_\pm$ \cite{Strominger:2013jfa}.  Independently, it accounts for the reduction of the number of degrees of freedom of soft gravitons in the semi-classical computation of on-shell gravity amplitudes which is necessary to prove Weinberg's soft graviton theorem \cite{Weinberg:1965nx} from the Ward identities associated with BMS supertranslations \cite{Strominger:2013jfa,He:2014laa}. Since the boundary conditions \eqref{BC1} are rooted in the nature of radiation, we now make the assumption that the boundary condition \eqref{BC1} are still valid in the presence of a final massive stationary state in the bulk spacetime. 

The solution to the boundary condition \eqref{BC1} is $C_{zz} = -2 D_z^2 C$ at both $\mathfrak I^+_\pm$ \cite{Strominger:2013jfa} where $C$ was called the Goldstone boson for spontaneous breaking of supertranslation invariance \cite{He:2014laa}. We will call it the asymptotic supertranslation field. It will play a crucial role in the following. In the final state, Einstein's equation $G_{uz} = 8 \pi G T_{uz}$ implies $\p_z m_B = -\frac{1}{4}\p_z (D_z^2 C^{zz} - D_{\bar z}^2 C^{\bar z \bar z})$ after assuming that the matter stress-tensor falls off sufficiently fast. The boundary condition \eqref{BC1}  then implies that the Bondi mass is a constant independent of $z,\bar z$ in the final state as claimed earlier.

Let us define the differential operator
\bea
\mathcal D = \frac{1}{4} D^2 (D^2 + 2).  \label{DD}
\eea
We can rewrite \eqref{c3} as 
\bea
M_{tot} -M_{stat} =  \left[\mathcal D C \right]^{\infty}_{u=-\infty} +  \int_{-\infty}^{\infty} du T_{uu}  . \label{c3bis}
\eea
We used the identity $\mathcal D C = (\gamma^{z \bar z})^2 D_z^2 D_{\bar z}^2C$. This conservation equation at $\mathfrak I^+$ is at the origin of gravitational memory effects \cite{Strominger:2014pwa,Pasterski:2015tva}.

We can also integrate \eqref{c2} between $v = -\infty$ and $v = +\infty$. The initial Bondi mass is denoted as $M_{in}$.  The final Bondi mass is $M_{tot}$ since it has to agree with the initial Bondi mass defined at $\mathfrak I^+_-$. This is the junction condition of the Bondi mass at spatial infinity  \cite{Strominger:2013jfa}. We impose the analogous boundary condition \eqref{BC1} at $\mathfrak I^-_\pm$, which allows to identify $C_{ww} = -2 D_w^2 C$  at both $\mathfrak I^-_\pm$. Therefore,
\bea
M_{tot} - M_{in} = -\left[ \mathcal D C \right]^{\infty}_{v=-\infty} +  \int_{-\infty}^{\infty} dv T_{vv} \label{c4}
\eea

The antipodal map leaves the operator $D^2$ and therefore $\mathcal D$ invariant. We can now subtract \eqref{c3bis} and \eqref{c4} to obtain 
\bea
\mathcal D \left( C_{stat}(z,\bar z) - \Delta C_\infty  - C_{in}(w,\bar w) \right) =  M_{in} - M_{stat}  +  \int_{-\infty}^{\infty} dv T_{vv}(w,\bar w)  -  \int_{-\infty}^{\infty} du T_{uu}(z,\bar z) \label{cons}
\eea
where $C_{stat}(z,\bar z) = \lim_{u \rightarrow \infty} C(u,z,\bar z)$, $C_{in}(w,\bar w) =\lim_{v \rightarrow -\infty} C(v,w,\bar w)$  and $\Delta C_\infty = \lim_{u \rightarrow -\infty} C(u,z,\bar z) - \lim_{v \rightarrow +\infty} C(v,w,\bar w)$. Let us pause to interpret this equation. The difference of the final asymptotic supertranslation field at angle $(z,\bar z)$  and the initial asymptotic supertranslation field at the antipodal angle $(w,\bar w)$ minus a boundary term at spatial infinity is dictated by the total energy flux and initial mass coming in minus the final mass and energy flux going out.\footnote{Canonical fields defined at spatial infinity are usually assumed to be even under parity ($\theta \rightarrow \pi - \theta$, $\phi \rightarrow \phi+\pi$) \cite{Regge:1974zd}. If the net stress-tensor flux in \eqref{cons} contains a parity odd piece, the asymptotic supertranslation field $C_{stat} - \Delta C_\infty - C_{in}$ will not be parity even and supertranslations might be broken by a classical anomaly \cite{Compere:2011ve}.}

 Let us now discuss the boundary term at spatial infinity $\Delta C_\infty $ in more detail. For small non-linear deviations of Minkowski spacetime, the junction conditions at spatial infinity consistent with Poincar\'e invariance was shown to be consistent with \cite{Strominger:2013jfa}
\bea
\lim_{u \rightarrow -\infty } C(u, z,\bar z) = -\lim_{v \rightarrow +\infty } C(v, w,\bar w) .\label{junc2}
\eea
If one starts with a configuration with no initial supertranslation field $C_{in}=0$ and generic incoming radiation, the conservation equation \eqref{c4} implies that $\lim_{v \rightarrow +\infty } C(v, w,\bar w) \neq 0$ and therefore $\Delta C_\infty \neq 0$ as a consequence of \eqref{junc2}. Now, we are interested in black hole collapse with no outgoing radiation, which is a different setting without limit to the scattering setting in Minkowski spacetime. In this setting, we will assume that the boundary term at spatial infinity is zero, $\Delta C_\infty = 0$. This ad hoc assumption does not contradict anything we know but it will need to be assessed by different considerations. It constitutes the main caveat of the remaining of this section. 

Let us now discuss the collapse of a spherically symmetric null shell. The process is described by the Vaidya metric
\bea
ds^2 = - (1- \frac{2 M \Theta(v)}{r} )dv^2 + 2 dv dr + r^2 d\Omega^2. 
\eea
The initial state is the global vacuum state with $C_{in}=0$ and $M_{in} = 0$. The stress-tensor is purely ingoing and given by $T^M_{\mu\nu} = \frac{M \delta (v) }{4 \pi r^2}\delta_\mu^v \delta_\nu^v$. There is no outgoing radiation so $T_{uu} = 0$. The final state is a black hole of mass $M_{stat} = M$.  In that very particular case and under our assumption $\Delta C_\infty = 0$, the conservation equation \eqref{cons} implies that $\mathcal D C_{stat} = 0$. The only smooth zero modes of the operator $\mathcal D$ are the four lowest spherical harmonics. Such harmonics are fixed by defining the center-of-mass frame which is centered at $r=0$. We deduce that $C_{stat} = 0$. There is therefore no non-trivial asymptotic supertranslation field  in the collapse of a spherically symmetric null shell. The collapse leads to the Schwarzschild black hole. 

Let us now take a non-spherical shell. Its energy is instead
\bea
T_{vv} = \left( \frac{M P^{in}(w,\bar w)}{4 \pi r^2} +O(r^{-3}) \right)\delta(v). 
\eea
The profile of the energy density on the spherical shell admits the harmonic decomposition $M P^{in}(w,\bar w) = M+M \sum_{l \geq 1,m} P_{l,m}Y_{l,m}$. We normalized the zero mode in order to uniquely define the mass. 

The null energy condition requires that $T_{vv} \geq  0$ at all angles. This implies $P^{in}(w,\bar w) \geq 0$ and it thereby constraints the coefficients of the higher harmonics $P_{l,m}$ as
\bea
\sum_{l \geq 1,m} P_{l,m}Y_{l,m}(\theta,\phi)  \geq -1.\label{constr}
\eea
For simplicity, we assume that the initial state is $M_{in} = C_{in} = 0$\footnote{The case $C_{in} \neq 0$ could be attributed to the presence of initial matter carrying the supertranslation field  or initial cosmological defects \cite{Compere:2016jwb}. We assume in this paper that the asymptotically flat spacetime patch under consideration admits no cosmological defect of the type discussed in  \cite{Compere:2016jwb}.}. In general, the non-linearities of Einstein's equations will lead to gravitational wave emission at $\mathfrak I^+$. Let us denote by $\int_{-\infty}^\infty du T_{uu} = \left( \frac{M P^{out}(z,\bar z)}{4 \pi r^2} +O(r^{-3}) \right)$ the total leading order outgoing energy flux profile. The conservation equation \eqref{cons} reads as 
\bea
\mathcal D C_{stat} =  M (P^{in}(w,\bar w) - P^{out}(z,\bar z) ) - M_{stat}. \label{eq:7}
\eea
In the ideal case where the outgoing radiation is negligeable, $P^{out} = 0$, the final mass is the initial mass, $M_{stat} = M$, and the zeroth spherical harmonic in $P^{in}$ cancels out with the last term in \eqref{eq:7}. Now, the presence of higher harmonics in $T_{vv}$ implies that there is a non-trivial profile for the supertranslation field  final state. After using  the properties of spherical harmonics $D^2 Y_{l,m}= - l (l+1)Y_{l,m}$ and $Y_{l,m}(\pi -\theta,\phi+\pi)=(-1)^l Y_{l,m}(\theta,\phi)$, we find in that ideal case without outgoing radiation,
\bea
C_{stat} =\sum_{l \leq 1,m} C_{l,m}^{(0)} Y_{l,m} +  M \sum_{l \geq 2,m}(-1)^l \frac{4}{(l-1)l(l+1)(l+2)} P_{l,m}Y_{l,m}. \label{rel}
\eea
The coefficients $C^{(0)}_{l,m}$ label the four zero modes of the differential operator $\mathcal D$ which are the 4 lowest spherical harmonics. They correspond to the center-of-mass of the system. The null energy condition constraint \eqref{constr} leads to non-trivial constraints on $C_{stat}$. 

Let us consider two simple toy models which we will use in the next section. If the non-sphericity of the ingoing null shell is only modelled by the $l=2$ $m=0$ spherical harmonic and in the center-of-mass frame, the final supertranslation field is
\bea
C_{stat} = \alpha \frac{M}{6} (3 \cos^2\theta - 1), \qquad -\frac{1}{2} \leq \alpha \leq 1\label{toy1}
\eea
where the amplitude $\alpha$ has been  constrained by the null energy condition \eqref{constr}. In the case where $P^{in}$ is 1 plus a combination of $l=2$ $m=\pm 1$ harmonics, we instead have 
\bea
C_{stat} = \alpha \frac{M}{6} \sin 2\theta \cos (\phi + \delta), \qquad -1 \leq \alpha \leq 1,\qquad \delta \in \mathbb R. \label{toy2}
\eea

As a summary, the spherical collapse of a non-spherical null shell leads to a final state which admits a non-trivial supertranslation field which can be determined from the 
``angle-dependent energy balance conservation law'' \eqref{cons}. The collapse of a spherically symmetric null shell is analytic and allows to identify that the final state supertranslation field vanishes. In more general cases, detailed numerical simulations would be necessary to compute for each individual collapse and subsequent evolution of the system the value of the final state asymptotic supertranslation field. 

The Schwarzschild black hole is therefore an extremely fine-tuned final state of gravitational collapse. It admits no supertranslation field. It only applies to black holes that formed in a spherically symmetric fashion such as the Vaidya spacetime. We now turn to our second question: what is the metric of a \emph{generic} classical final state of gravitational collapse?

\section{Supertranslation-dependent final state metric}
\label{sec:cg}

The classical final state after gravitational collapse is by definition a stationary spacetime. Uniqueness theorems imply that the spacetime metric is diffeomorphic to the Kerr metric after an assumption of analyticity or small deviation from the Kerr metric \cite{Hawking:1973uf,Carter:1971zc,Robinson:1975bv,Chrusciel:2008js,Alexakis:2009ch}. In this paper, for technical simplicity, we assume that the spacetime is static (the metric is invariant under time reversal). We will construct a final state diffeomorphic to the Schwarzschild metric. The stationary case which takes the angular momentum into account will be considered elsewhere. 
 
The task at hand is to understand the group of diffeomorphisms preserving the asymptotically flat boundary conditions and generating the asymptotic supertranslation field. Part of the group of diffeomorphisms is pure gauge in the sense that it corresponds to changing coordinates without changing any physics. Diffeomorphisms changing the asymptotic supertranslation field are physical in the sense that they change the physical charges of the system, namely the superrotation charges as we will explicitly show following \cite{Compere:2016jwb}. In the vacuum case, it was shown in \cite{Compere:2016jwb} that superrotation charges are also defined in the bulk of the spacetime, which implies that the diffeomorphism which generates the supertranslation field is singular in the bulk. Otherwise, one could deform the sphere of integration from infinity to a bulk point and the charge would be zero, but it is not. So, the supertranslation-generating diffeomorphisms are necessarily singular in the bulk. In the case of a black hole formed by collapse, the singularities have to be hidden behind an event horizon if the weak cosmic censorship conjecture \cite{Penrose:1969pc,Hawking:1976ra} holds. We emphasize that even if singularities are introduced by a supertranslation diffeomorphism, they cannot be considered coordinate singularities or pure gauge since the physical canonical charges (the superrotation charges in this case) are fixed for a given configuration and therefore cannot be gauged away\footnote{A similar situation happens in the $AdS_3$ case in the presence of a non-trivial boundary stress-tensor where the Virasoro charges are non-trivial \cite{Brown:1986nw,Banados:1998gg,Compere:2015knw}.}. As we will describe, the bounds derived from the null energy condition in the last section will ensure that such singularities are hidden by an infinite redshift surface. Our construction will therefore be compatible with the cosmic censorship conjecture.

In this section we will describe the simplest construction of a final state with an arbitrary fixed asymptotic supertranslation field profile $C(z,\bar z)$. It will be constructed by a specific large diffeomorphism applied to the Schwarzschild metric which we could construct explicitly. We emphasize that applying a large finite diffeomorphism is not a physical process: it is a convenient solution generating technique which allows to get the final state of gravitational collapse with a non-trivial asymptotic supertranslation field.

\subsection{Metric from the supertranslation diffeomorphism}

We start with the Schwarzschild metric in Boyer-Linquist coordinates $(t_s,r_s,\theta_s,\phi_s)$,
\bea
ds^2 = -(1- \frac{2M}{r_s})dt_s^2 + \frac{1}{1-\frac{2M}{r_s}}dr_s^2 + r_s^2 d\Omega_s^2.
\eea
Here, $d\Omega^2  = d\theta_s^2 + \sin^2\theta_s d\phi_s^2 = 2\gamma_{z_s \bar z_s}dz_s d\bar z_s = \gamma_{AB}dz_s^A dz_s^B$ is the unit metric on the round sphere. The event Killing horizon is located at the infinite redshift surface $r_s = 2M$. It is well-known that spatial sections of the Schwarzschild black hole are conformally flat. This feature can be made manifest using the isotropic coordinate system $(t_s,\rho_s,\theta_s,\phi_s)$,
\bea
ds^2 =- \frac{\left( 1-\frac{M}{2\rho_s}\right)^2}{\left( 1+\frac{M}{2\rho_s}\right)^2}dt_s^2 + \left( 1+\frac{M}{2\rho_s}\right)^4 (d\rho_s^2 + \rho_s^2 d\Omega_s^2). 
\eea
The radial coordinates $r_s$ and $\rho_s$ are related as $r_s=\rho_s ( 1+\frac{M}{2\rho_s})^2$. The isotropic radial coordinate interpolates between $\infty$ at spatial infinity and $\frac{M}{2}$ at the horizon.

We now consider the solution generating technique which consists in applying a large coordinate transformation which introduces the supertranslation field  in the metric. We denote the final coordinates as $(t,\rho,\theta,\phi)$. The supertranslation field  is defined as an arbitrary function on the sphere $C(\theta,\phi)$ which transforms under supertranslations as 
\bea
\delta_T C(\theta,\phi) = T(\theta, \phi). \label{deltaC}
\eea
We call the coordinate transformation large in order to distinguish it from a gauge transformation which by definition leaves all physical quantities invariant. Here, it will change the asymptotic supertranslation field $C(\theta,\phi)$ which is physically fixed by the details of the collapse as we discussed earlier. 

After a long computation which we outline in Appendix A, we found such a diffeomorphism in static gauge where $g_{\rho\theta} = g_{\rho \phi} = 0$. In accordance with the time reversal $Z_2$ symmetry, it leaves the time component of the metric unchanged and it only transforms non-trivially the three-dimensional spatial metric $ds^2_{(3)}$ which can be rewritten as
\bea\label{trands3}
ds^2_{(3)} \equiv d\rho_s^2 + \rho_s^2 d\Omega_s^2 = d\rho^2 + \left( ((\rho- E)^2 + U)\gamma_{AB} + (\rho - E) C_{AB} \right)dz^Adz^B.
\eea
Here the auxiliary quantities $C_{AB}$, $U$ and $E$ are defined in terms of $C$ as
\bea
C_{AB}(\theta,\phi) &\equiv & -(2D_AD_B-\gamma_{AB} D^2)C,\nn\\ 
U(\theta,\phi) &\equiv & \frac{1}{8}C_{AB}C^{AB},\label{aux} \\
E(\theta,\phi) &\equiv & \frac{1}{2}D^2C +C - C_{(0,0)}. \nn
\eea
All uppercase indices $A,B,\dots$ are raised with $\gamma^{AB}$. The diffeomorphism reads as 
\bea
t_s &=& t + C_{(0,0)}, \nn \\
\rho_s &=& \sqrt{(\rho - C + C_{(0,0)})^2 + D_A C D^A C}, \label{diffeo}\\
z_s &=& \frac{(z- \bar z^{-1}) (\rho - C + C_{(0,0)}) + (z + \bar z^{-1}) (\rho_s - z \p_z C - \bar z \p_{\bar z} C)}{2(\rho - C +C_{(0,0)}) + (1+z \bar z)(\bar z \p_{\bar z} C - \bar z^{-1} \p_z C)}.\nn
\eea
We define the final spherical coordinates $(\theta,\phi)$ by the stereographic map $z = e^{i \phi} \cot\frac{\theta}{2}$. Here $C_{(0,0)}$ denotes the lowest spherical harmonic mode of $C$. A time translation is generated by $C = C_{(0,0)}= 1$. A spatial translation is generated by 
\bea
C_{\text{translation}} = a_x \sin\theta \cos\phi + a_y \sin\theta \sin\phi + a_z \cos\theta\label{Ctran}
\eea
which is a linear combination of the $l=1$ spherical harmonics. The transformation \eqref{diffeo} then precisely coincides with the transformation law of the spherical coordinate system centered at the origin to a new spherical coordinate system centered at the translated origin. The transformation law of the radius follows from Pythagoras' theorem. For a generic supertranslation, the transformation is still given by \eqref{diffeo} but where $C$ is now an arbitrary combination of higher spherical harmonics. We will therefore refer to the transformation rule of the radius \eqref{diffeo} as the \emph{supertranslation Pythagorian rule}. All supertranslations except the lowest harmonic mode are purely spatial.

The spatial part of the diffeomorphism \eqref{diffeo} takes a much simpler form when it transforms the original flat metric in Cartesian coordinates $(x_s,y_s,z_s)$ to the final one in spherical coordinates $(\rho,z^A)$ with $z^A=\theta,\phi$ as
\bea
ds^2_{(3)}  = dx_s^2 + dy_s^2+dz_s^2 = d\rho^2 + \left( ((\rho- E)^2 + U)\gamma_{AB} + (\rho - E) C_{AB} \right)dz^Adz^B. \label{eq:89}
\eea
The coordinates then transform as 
\begin{align}
x_s&=(\rho-C+C_{(0,0)}) \sin\theta \cos\phi+\csc\theta\sin\phi\partial_{\phi}C-\cos\theta\cos\phi\partial_{\theta}C,\nn\\
y_s&=(\rho-C+C_{(0,0)})\sin\theta\sin\phi-\csc\theta\cos\phi\partial_{\phi}C-\cos\theta\sin\phi\partial_{\theta}C,\\
z_s&=(\rho-C+C_{(0,0)})\cos\theta+\sin\theta\partial_{\theta}C. \nn 
\end{align}
After defining $x = \rho \sin\theta \cos\phi$, $y = \rho\sin\theta \sin\phi$, $z = \rho \cos\theta$, a spatial translation generated by \eqref{Ctran} leads to the diffeomorphism $x_s = x - a_x$, $y_s = y -a_y$, $z_s = z - a_z$, as expected.

After applying this diffeomorphism, the static metric with a non-trivial asymptotic supertranslation field turned on is 
\bea
ds^2 =- \frac{\left( 1-\frac{M}{2\rho_s}\right)^2}{\left( 1+\frac{M}{2\rho_s}\right)^2}dt^2 + \left( 1+\frac{M}{2\rho_s}\right)^4 \Big( d\rho^2 + \left( ((\rho- E)^2 + U)\gamma_{AB} + (\rho - E) C_{AB} \right)dz^Adz^B \Big)\label{cgm}
\eea
where auxiliary quantities are defined in \eqref{aux} and $\rho_s = \sqrt{(\rho - C + C_{(0,0)})^2 + D_A C D^A C}$ as given in \eqref{diffeo}. This is the Schwarzschild metric equipped with a supertranslation field. 

The generator of an arbitrary supertranslation which preserves the form of the metric \eqref{cgm} in static gauge takes the form 
\be
\xi^{(stat)}_T=T_{(0,0)} \partial_t-(T(\theta,\phi) -T_{(0,0)}+O(\rho^{-1}))\partial_{\rho}- (\frac{D^A T(\theta ,\phi )}{\rho} +O(\rho^{-2})) \partial_A.\label{st}
\ee
Here $T_{(0,0)}$ is the $l=m=0$ harmonic of the generic regular function $T(\theta,\phi)$ which generates time shifts. The spatial translations are generated by the $l=1$ harmonics of $T$ and the other supertranslations are generated by the higher harmonics of $T$.  One can adjust the subleading terms in \eqref{st} such that the metric $g_{\mu\nu}(C ; M)$ \eqref{cgm}  exactly transforms under infinitesimal supertranslations as 
\bea
\mathcal L_{\xi^{(stat)}_T} g_{\mu\nu}(C ; M) = \lim_{\eps \rightarrow 0} \frac{g_{\mu\nu}(C+ \epsilon \, T ; M) -g_{\mu\nu}(C ; M) }{\epsilon } \equiv \delta_{T} g_{\mu\nu}(C ; M) .
\eea
Here, the variation $\delta_{T}$ acts on the field $C$ and its $p$-th derivative, $p=1,2,\dots$ as a derivative operator contracted with the $p$-th derivative of $\delta_{T} C(\theta,\phi) = T(\theta,\phi)$ as defined in \eqref{deltaC}. After some algebra, the exact generator of supertranslations in static gauge is found to be 
\be
\xi^{(stat)}_T=T_{(0,0)}\partial_t-(T-T_{(0,0)})\partial_{\rho}+\frac{C^{AB}D_BT-2D^AT(\rho-\frac{1}{2}(D^2+2)(C-C_{0,0}))}{2((\rho-\frac{1}{2}(D^2+2)(C-C_{0,0}))^2-U)}\partial_A. \label{xisM}
\ee
These generators coincide with the ones obtained in the massless case $M=0$ \eqref{xis} thanks to the remarkable property
\bea
\xi^{(stat) \mu }_T \frac{\p}{\p x^\mu_{(stat)}} \rho_s(\rho,\theta,\phi) = \delta_{T} \rho_s(C)
\eea
where $x^\mu_{(stat)} = (t,\rho,\theta,\phi)$. 

The generators \eqref{xisM} exactly commute under the adjusted bracket defined in \cite{Barnich:2010eb}
\be
[\xi_1,\xi_2]_{ad} \equiv [\xi_1,\xi_2]-\delta_{\xi_1}\xi_2+\delta_{\xi_2}\xi_1.
\ee
As a consequence of these vanishing commutation relations, the supertranslations act everywhere in the bulk spacetime described by the metric \eqref{cgm}. This extends in the bulk the observations made at null infinity in \cite{Barnich:2010eb}. In group theory language, the metric \eqref{cgm} describes the orbit of the Schwarzschild black hole under the supertranslation group. 

In the vacuum case $M=0$, the generator in static gauge \eqref{xisM} can be related to the generator in $BMS_\pm$ gauge which naturally acts at future ($+$) or past ($-$) null infinity as (see Appendix \ref{Diff1})
\bea
\xi^{(BMS_\pm)}_T = \xi^{(stat)}_T - \delta_{T} x_{(BMS_\pm)}^\mu \frac{\p}{\p x^\mu_{(BMS_\pm)}}\label{relas}
\eea
where $x^\mu_{(BMS_+)} = (u,r,z,\bar z)$ and $ x^\mu_{(BMS_-)} = (v,r,w,\bar w)$. We expect that this relationship extends straightforwardly in the presence of mass $M$ but we didn't check it explicitly. The relationship \eqref{relas} allows to link the BMS symmetries at null infinity with the BMS symmetries at spatial infinity simply because the symmetries are defined in the entire bulk spacetime. This completes previous arguments on the existence of such a map \cite{Ashtekar:1978zz}.

\subsection{Kinematical properties}

In the spherical symmetric collapse of a null shell, we showed that $C = 0$ and the metric is the Schwarzschild metric with an event Killing horizon at $\rho = \rho_s = \frac{M}{2}$. The final state of a general collapse is not described by the Schwarzschild metric since there is no gauge transformation that allows to switch off the asymptotic supertranslation field $C$. At large $\rho$, one has $\rho \sim \rho_s$ and the metric is asymptotically flat but it finitely deviates from the Schwarzschild metric at finite $\rho$. This leads to new features. 

The first subleading order in $\rho$ from the Schwarzschild metric is fixed by the linearized BMS asymptotic symmetry structure. There are two effects at first subleading order. First, the spherical metric $\rho^2\gamma_{AB} dz^A dz^B = \rho^2 d\Omega^2$ is deformed as
\bea
\rho^2d \Omega^2 \rightarrow \rho^2d\Omega^2 + \rho (C_{AB}dz^A dz^B - 2 E  d\Omega^2)+O(\rho^0)
\eea
Second, the Schwarzschild radius $\rho_s$ is shifted as $\rho_s = \rho - (C - C_{(0,0)})+O(\rho^{-1})$.\footnote{The sphere deformation only depends upon the $l \geq 2$ harmonics of the supertranslation field but the shift of the radius $\rho_s$ also depends upon the $l=1$ harmonics attributed to the change of center-of-mass frame. Indeed, $C_{AB}$ is independent of the $l \leq 1$ harmonics in $C$ and the operator $D^2+2$ annihilates the $l=1$ spherical harmonics of $C$.} At subsequent subleading orders, the structure becomes non-linear in the supertranslation field. 

Before interpretating the solution further, it is necessary to further understand the diffeomorphism \eqref{diffeo}. At this stage, it might be useful for the reader to first read the toy model of vacuum 3-dimensional Einstein gravity described in Appendix \ref{threed} which possesses similar qualitative features but where the algebra is easier. 

The determinant of the 3-dimensional metric \eqref{trands3} reads in the new coordinates as
\bea
\sqrt{\text{Det} g^{(3)}_{ab}} = \sqrt{\text{Det} \gamma_{AB}} | (\rho - E)^2 - U |. 
\eea
We deduce that the Jacobian of the diffeomorphism \eqref{diffeo} is $ | (\rho - E)^2 - U |$. This Jacobian vanishes at a specific locus which we will denote as $\rho = \rho_{SH}(z,\bar z)$ where\footnote{The locus $\rho_s = 0$ is also a coordinate singularity of the original coordinates $(\rho_s,\theta_s,\phi_s)$ but is lies beyond the range of $\frac{M}{2} \leq \rho_s < \infty$ so we can ignore it.}
\bea\label{defrhoD}
\rho_{SH}(z,\bar z) =  E + \sqrt{U} =  \frac{1}{2}D^2C +C - C_{(0,0)} + \frac{1}{2\sqrt{2}} \sqrt{C_{AB}C^{AB}}. 
\eea
Indeed, one can check that the determinant of the spherical part of the metric \eqref{trands3} at fixed $\rho = \rho_{SH}$ vanishes
\bea
\text{Det}\left( 2 U \gamma_{AB} + \sqrt{U} C_{AB} \right) = 0, \label{van}
\eea
after using the property $\gamma^{AB}C_{AB} = 0$ and the definition of $U$.

\begin{figure}[!tbh]
\centering
\includegraphics[scale=0.50]{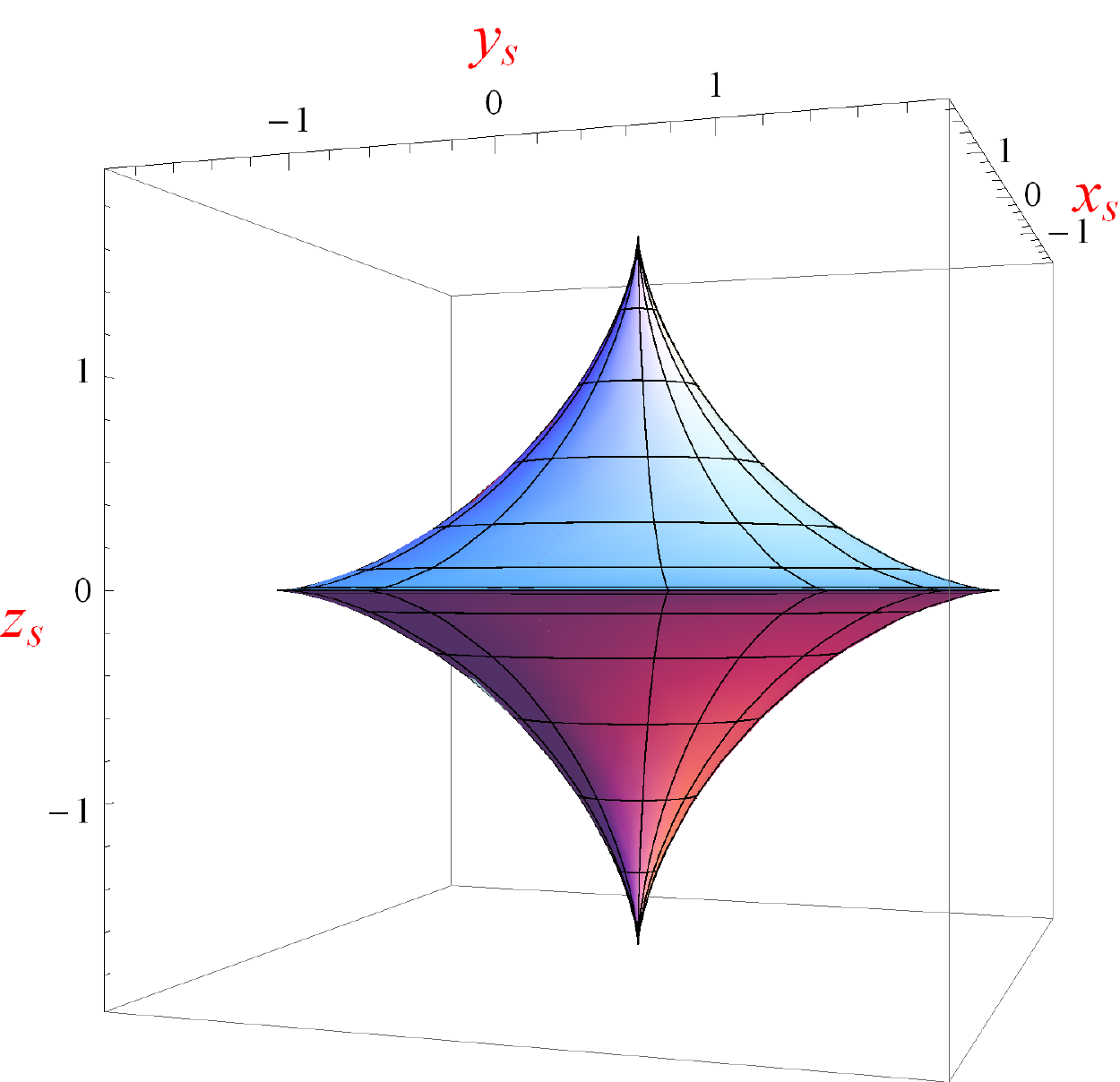}
\caption{Isometric embedding of the supertranslation horizon $\rho = \rho_{SH}(\theta,\phi)$ in 3-dimensional Euclidean space $ds_{(3)}^2 = dx_s^2+dy_s^2+dz_s^2$ as defined from \eqref{eq:89} and \eqref{defrhoD}. The supertranslation field is chosen to be the lowest non-trivial axisymmetric $l=2$, $m=0$ spherical harmonic $C(\theta,\phi)=Y_{2,0}(\theta,\phi)$.}
\label{fig4}
\end{figure}

\begin{figure}[!bth]
\begin{minipage}{0.45\textwidth}
\centering
\includegraphics[scale=0.47]{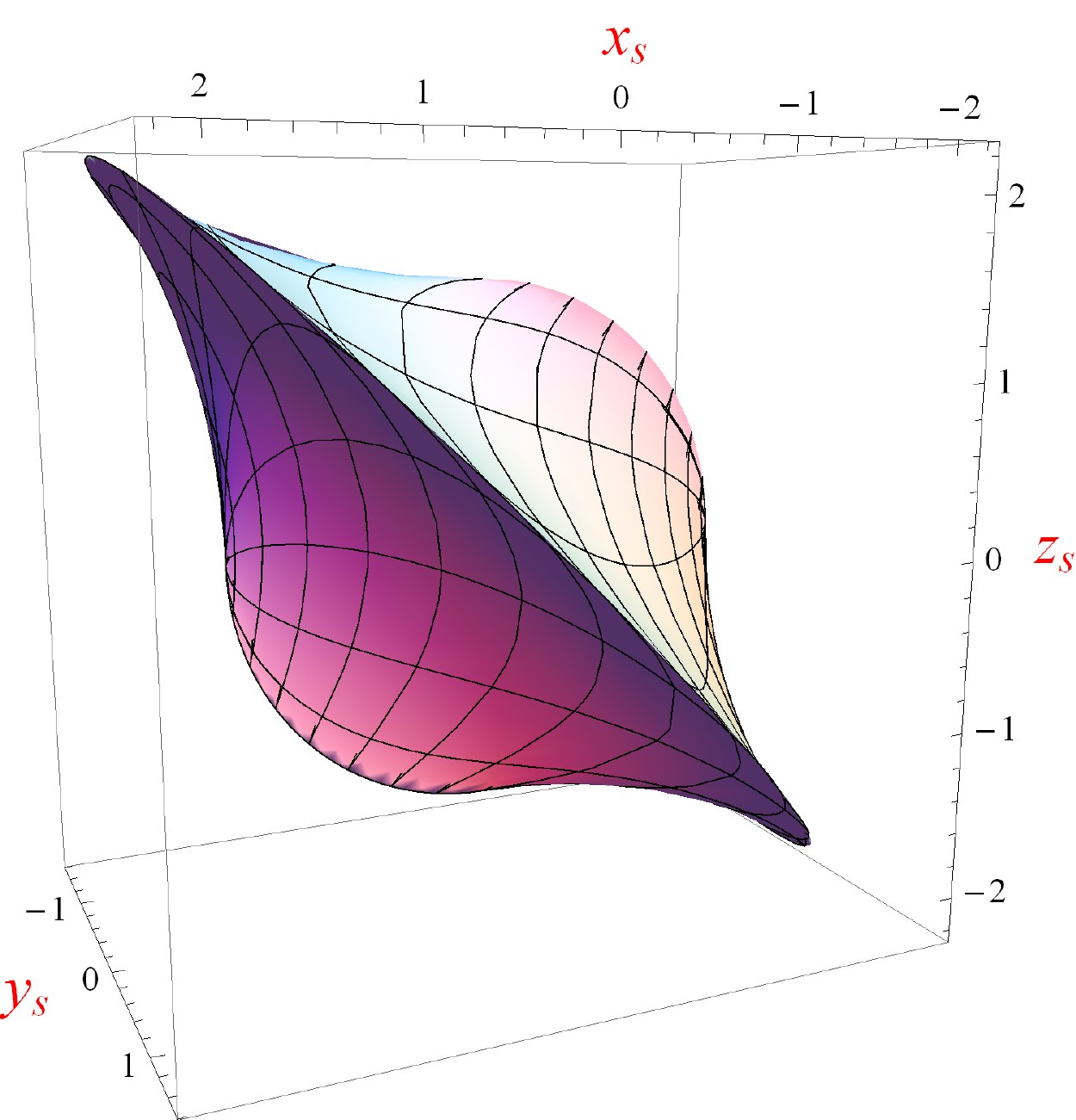}
\caption{Isometric embedding of the supertranslation horizon in the case $C(\theta,\phi)=Y_{2,1}(\theta,\phi)-Y_{2,-1}(\theta,\phi)$.}
\label{fig5}
\end{minipage}\hfill
\begin{minipage}{0.45\textwidth}
\centering
\includegraphics[scale=0.50]{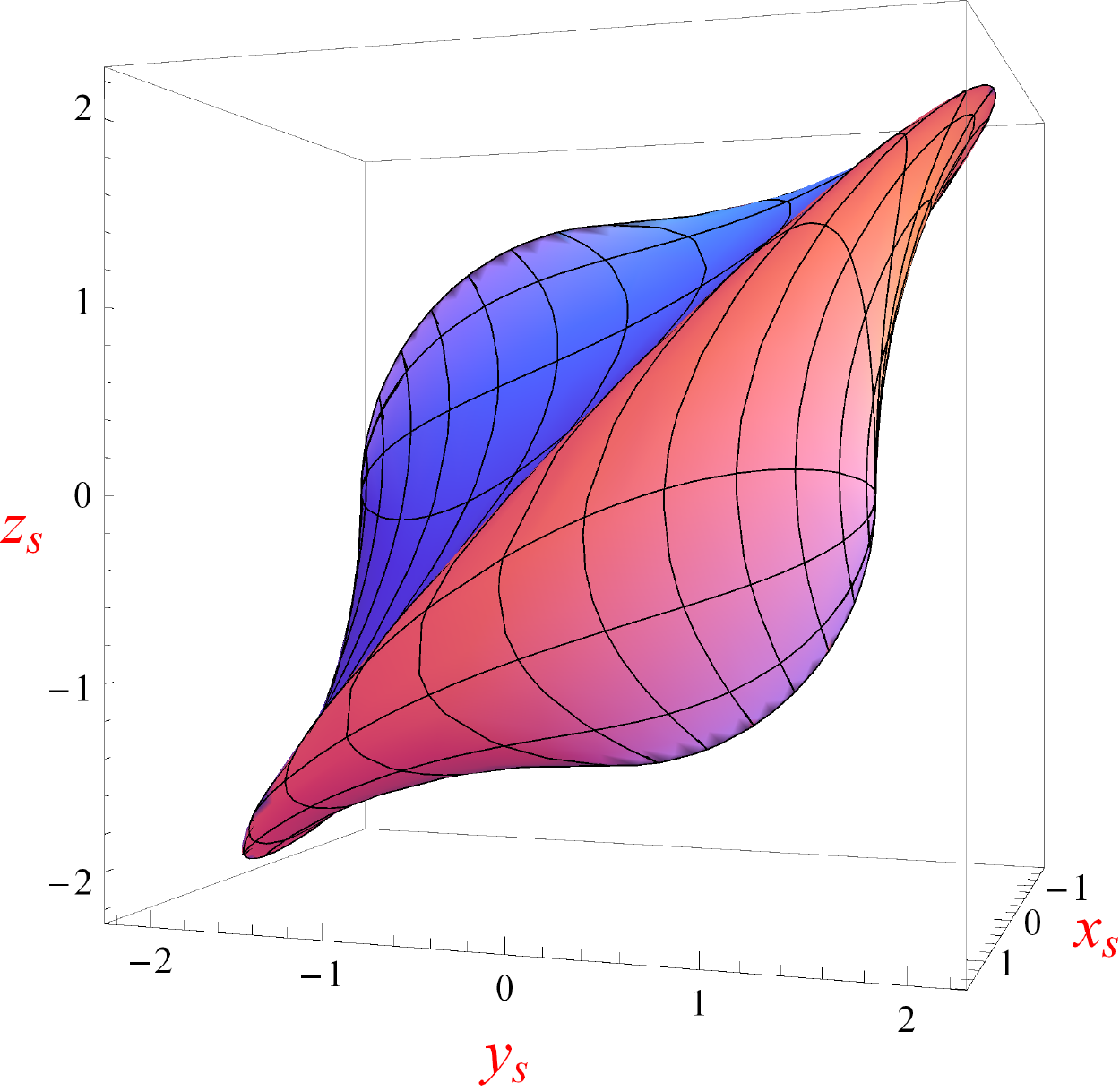}
\caption{Isometric embedding of the supertranslation horizon in the case $C(\theta,\phi)=-i(Y_{2,1}(\theta,\phi)+Y_{2,-1}(\theta,\phi))$.}
\label{fig6}
\end{minipage}
\end{figure}

The induced metric $\gamma^{SH}_{AB} $ on the surface $\rho = \rho_{SH}(z,\bar z)+\eps$ in the limit $\eps  \rightarrow 0$ is 
\bea
\gamma^{SH}_{AB} \equiv \widehat\gamma_{AB} +\p_A \rho_{SH} \p_B \rho_{SH}, \qquad \widehat\gamma_{AB} \equiv 2 U \gamma_{AB} + \sqrt{U} C_{AB}.
\eea
Its determinant is given by $\text{Det}(\gamma^{SH}_{AB} ) = \text{Det}(\widehat\gamma_{AB} )(1+\widehat\gamma^{AB}\p_A \rho_{SH} \p_B \rho_{SH})$ where $\widehat\gamma^{AB}$ is the inverse of $\widehat\gamma_{AB} $ which exists at finite $\eps$. Let us use spherical coordinates $z^A = (\theta,\phi)$.  The relation \eqref{van} implies at $\eps =0$, $(\widehat\gamma_{\theta \phi})^2 = \widehat\gamma_{\theta\theta} \widehat\gamma_{\phi\phi}$. The sign of  $\widehat\gamma_{\theta\phi}$ is equal to the sign of $C_{\theta\phi}$. 
The induced metric $\gamma^{SH}_{AB}$ admits the measure
\bea
\sqrt{\text{Det}(\gamma^{SH}_{AB} ) }= \Big| \sqrt{\widehat\gamma_{\theta \theta}} \p_{\phi} \rho_{SH}  - \text{sign}(\widehat\gamma_{\theta \phi}) \sqrt{\widehat\gamma_{\phi \phi}}  \p_\theta \rho_{SH} \Big|. \label{mea}
\eea
This measure is generically non-vanishing. We conclude that the locus $\rho = \rho_{SH}(\theta,\phi)$ is a 2-dimensional surface. This surface is related via a diffeomorphism to a finite region of Euclidean spacetime because $\rho_s(\rho_{SH}(\theta,\phi),\theta,\phi) \geq 0$. This inequality is a consequence of the supertranslation Pythagorian rule. Therefore, geodesics reach $\rho = \rho_{SH}(\theta,\phi)$ with finite affine parameter and such geodesics continue smoothly upon an infinitesimal increase of their affine parameter. We call this location the \emph{supertranslation horizon}. Since $\rho_{SH}$ is a smooth function on the sphere, there are at least two points where $\rho_{SH}$ is a local extremum, $\p_\theta \rho_{SH} = \p_\phi \rho_{SH} = 0$.  At these points, the induced metric on the surface $\rho = \rho_{SH}$ is degenerate. There might also be closed lines where the measure \eqref{mea} vanishes. One can check by explicit examples that the induced metric $\widehat\gamma_{AB}$ has line and points cusps, see Figures \ref{fig4}-\ref{fig5}-\ref{fig6}. 

The local geometry close to the supertranslation horizon can be investigated as follows. Let us consider the solid angle around the angle $(\theta_*,\phi_*)$: $\theta = \theta_* + \delta \theta$, $\phi = \phi_* + \delta \phi$ with $\delta\theta$, $\delta\phi$ small and $\rho_{SH} \leq \rho \leq \infty$ arbitrary. It is convenient to define the new radius $\rho_* = \rho - (E_*+\sqrt{U_*})$ which is $\rho$ shifted by a constant. Here $E_*$ and $U_*$ are the values of $E$ and $U$ at the angle $(\theta_*,\phi_*)$. The $\rho_*$ coordinate ranges as $0 \leq \rho_* < \infty$. In the vicinity of that angle the flat spatial metric $ds^2_{(3)}$ reads as 
\bea
ds^2_{(3)} = d\rho_*^2 + \left( (\rho_*^2+2\sqrt{U_*} \rho_* +2 U_*)\gamma^*_{AB} + (\rho_* + \sqrt{U_*}) C_{AB}^* \right) d\delta x^A d\delta x^B+O((\delta x^A)^3)
\eea
where $\gamma_{AB}^*$ and $C_{AB}^*$ are the value of $\gamma_{AB}$ and $C_{AB}$ at $x^A = (\theta_*,\phi_*)$ and $\delta x^A = (\delta \theta,\delta \phi)$. This metric can be written as
\bea
ds^2_{(3)} = d\rho_*^2 + \left( 1+\frac{\rho_*}{\sqrt{U_*}} \right) (d\delta z_*)^2 +\rho_*^2 \left(  (d\delta z_*)^2 +  (d\delta \phi_*)^2 \right) +O((\delta x^A)^3)\label{mls}
\eea
Indeed, the vanishing of the determinant \eqref{van} at $(\theta_*,\phi_*)$ implies that it exists a linear combination $\delta z_*$ of the coordinates $\delta \theta, \delta\phi$ such that $ (2 U_* \gamma^*_{AB} + \sqrt{U_*} C^*_{AB} )d\delta x^A d\delta x^B= (d\delta z_*)^2$. We can then define $d\delta \phi_*$ such that $\gamma^*_{AB}d\delta x^A d\delta x^B =  (d\delta z_*)^2 +  (d\delta \phi_*)^2$. \footnote{If $\theta_* = 0,\pi$ spherical coordinates cannot be used and $x^A$ then denotes Cartesian coordinates in a small patch around either the north or south pole. Our reasoning still holds with minor rewriting.} The metric \eqref{mls} is recognized as a solid angle at large $\rho_*$ which ends on a line segment spanned by $\delta z_*$ at $\rho_* = 0$. Close to $\rho_* = 0$ the metric describes a 2-dimensional solid angle in the $(\rho_*,\delta\phi_*)$ plane times a line direction $\delta z_*$ in cylindrical coordinates. Now, this solid angle naturally extends to a larger patch of 3-dimensional Euclidean space. One can just switch to local Cartesian coordinates and the solid angle is only a coordinate patch of the 3-dimensional space. We conclude that the supertranslation horizon $\rho = \rho_{SH}(\theta,\phi)$ is a smooth surface where space does not end.

Let us show that no curvature singularity lies inside of the coordinate patch. In the Schwarzschild spacetime, the locus where curvature invariants blow up lies at $r_s = 0$ in a coordinate patch beyond the patch covered by isotropic coordinates. Nevertheless, this divergence of curvature invariants can be detected by analytic continuation to $\rho_s = -\frac{M}{2}$ which is the zero of the analytically continued relation between $\rho_s$ and $r_s = \rho_s (1+\frac{M}{2\rho_s})^2$. Curvature invariants are invariant under diffeomorphisms. We can therefore detect the curvature singularities in the final metric \eqref{cgm} by asking whether or not the locus $\rho_s = -\frac{M}{2}$ lie within the bulk region bounded by the supertranslation horizon $\rho = \rho_{SH}$. The diffeomorphism \eqref{diffeo} implies that for $C \neq 0$, $\rho_s (\rho_{SH}(z,\bar z),z,\bar z) \geq  0$ because it is given by the square root of a sum of squares. Therefore, there is no curvature singularity in the coordinate patch.

Let us now discuss the relative location of the supertranslation horizon with respect to the infinite redshift surface or Killing horizon. The infinite redshift surface is located at isotropic radius $\rho = \rho_{H}(z,\bar z)$ which obeys $\rho_s(\rho_H) = \frac{M}{2}$. In order words it obeys
\bea
 (\rho_H(z,\bar z) - C + C_{(0,0)})^2 = \frac{M^2}{4} - D_A C D^A C. \label{rhoHeq}
\eea
For small $C \ll M$ with respect to the mass scale, $\rho_H \sim \frac{M}{2}$ and $\rho_H \gg \rho_{SH}$. The supertranslation horizon is therefore always hidden behind the infinite redshift surface where the coordinate patch ends. 

The existence of an infinite redshift surface depends upon the sign of $\frac{M^2}{4} - D_A C D^A C$. Let us now argue that if there is an angle $(\theta,\phi)$ such that 
\bea
\frac{M^2}{4} - D_A C D^A C < 0
\eea
then the cosmic censorship hypothesis would be violated.\footnote{As stated as the beginning of this section we recall that the diffeomorphism turning on the supertranslation field cannot be undone without changing the canonical superrotation charges and therefore the physical state. It might therefore lead to singularities which cannot be undone by any gauge transformation. Another independent necessary condition for cosmic censorship is $M>0$.}  For that angle, $\rho_H$ is imaginary, which implies that the would be infinite redshift surface lies beyond the coordinate patch bounded by the supertranslation horizon. Now, the location $\rho(\theta,\phi) = \rho_{MR}(\theta,\phi) \equiv C(\theta,\phi)$ is a location where $\rho_s(\rho_{MR}) = \sqrt{D^A C D_A C} > \frac{M}{2}$. At this angle, the location $\rho = \rho_{MR}$ is the minimum of $\rho_s(\rho)$. At radii smaller than $\rho_{MR}$, $\rho_s$ will increase again which implies that the redshift $1/\sqrt{-g_{tt}}$ will decrease again. The surface $\rho = \rho_{MR}$ is therefore a maximal finite redshift surface. It leads us to conclude that in that scenario there is no infinite redshift surface at all for this angle, and the singularity of either the supertranslation defect or the curvature singularity would be naked, which violates the cosmic censorship hypothesis. 

We are therefore led to conclude that the cosmic censorship hypothesis implies 
\bea
\frac{M^2}{4} - D_A C D^A C \geq 0.\label{bound}
\eea
Now, the value of the supertranslation field is fixed by the details of the collapse and this bound should therefore be confronted to prediction of the conservation equation \eqref{cons} which is constrained by the null energy condition. In the two toy models of non-spherical collapse where the supertranslation fields are \eqref{toy1} and \eqref{toy2}, we checked that the bound \eqref{bound} is obeyed. This is a non-trivial test of the validity of the bound \eqref{bound}. It is an interesting mathematical question to ask whether or not one could prove \eqref{bound} from the ideal model described in \eqref{constr}-\eqref{eq:7} which reduces in the simplest case to 
\bea
\mathcal D C = M \sum_{l \geq 2,m} P_{l,m} Y_{l,m}  \geq - M \label{ex}
\eea
Proving \eqref{bound} from \eqref{ex} would requires boundedness properties of the differential operator $\mathcal D$ defined in \eqref{DD}.

\begin{figure}[!hbt]
\begin{minipage}{0.45\textwidth}
\centering
\includegraphics[scale=0.70]{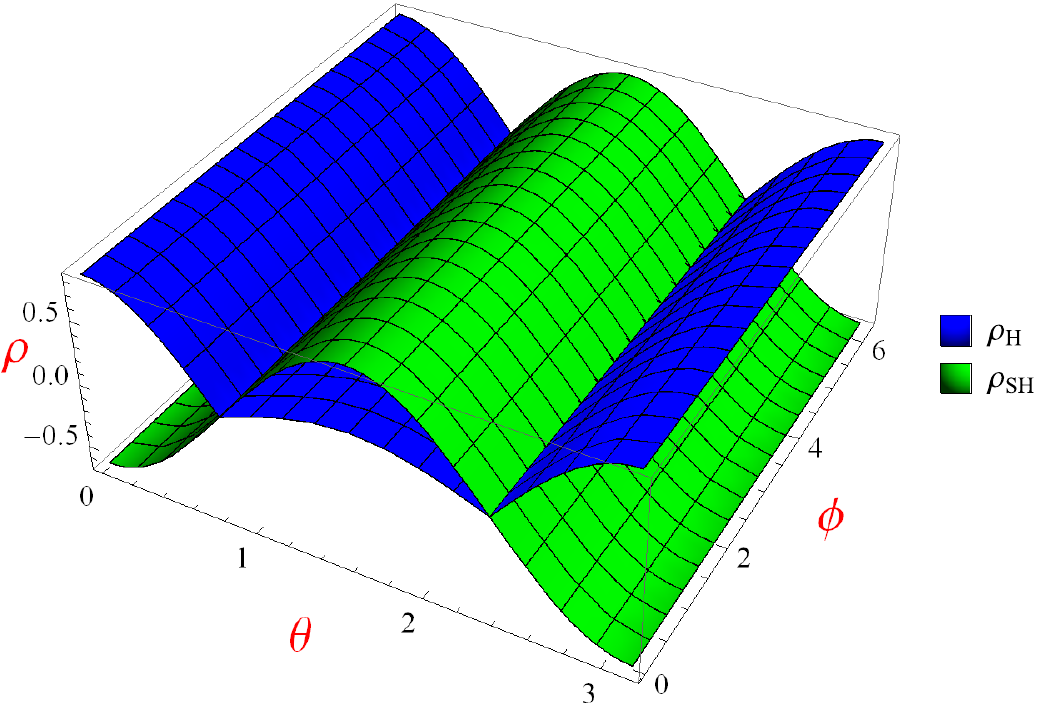}
\end{minipage}\hfill
\begin{minipage}{0.45\textwidth}
\centering
\includegraphics[scale=0.70]{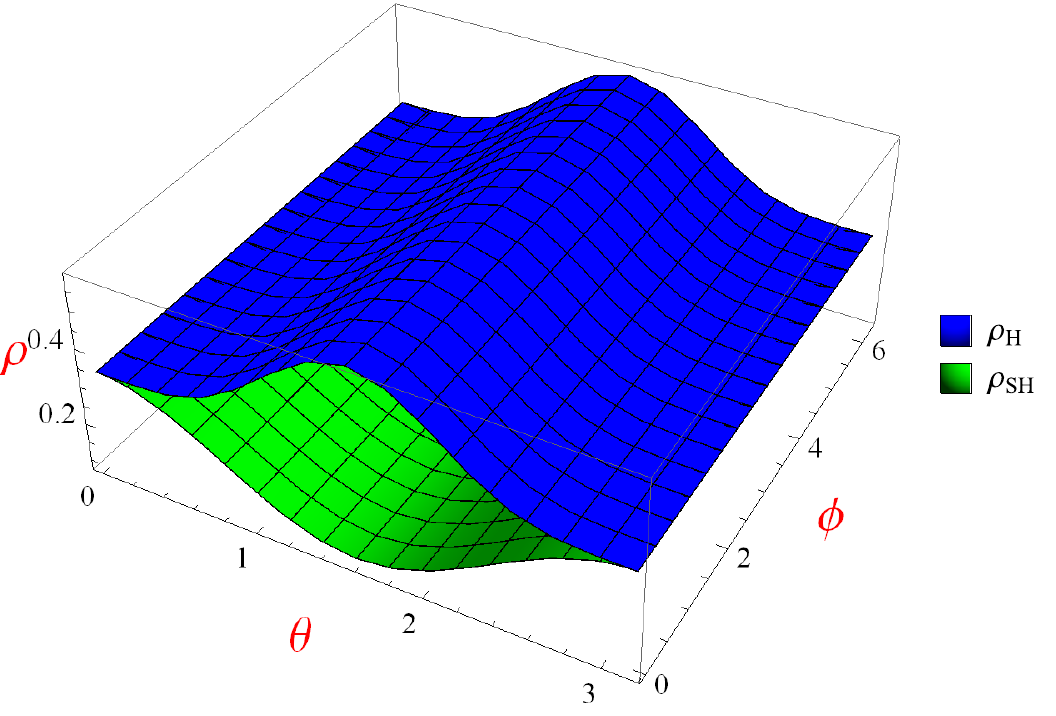}
\end{minipage}
\caption{Relative positions of the supertranslation horizon $\rho_{SH}$  and the Killing horizon $\rho_H$ as a function of the spherical angles $(\theta,\phi)$. On the left figure, $C(\theta,\phi)=\frac{M}{6}(3\cos^2\theta-1)$. This corresponds to the toy model \eqref{toy1} with $\alpha=1$. The Killing horizon is reached close to the north and south poles but the supertranslation horizon is reached first close to the equator.  On the right figure, $C(\theta,\phi)= -\frac{M}{12} (3\cos^2\theta-1)$. This corresponds to the toy model \eqref{toy1} with $\alpha=-\frac{1}{2}$. The supertranslation horizon is entirely shielded by the Killing horizon.}
\label{fig1}
\end{figure}

\begin{figure}[!hbt] 
\begin{minipage}{0.45\textwidth}
\centering
\includegraphics[scale=0.70]{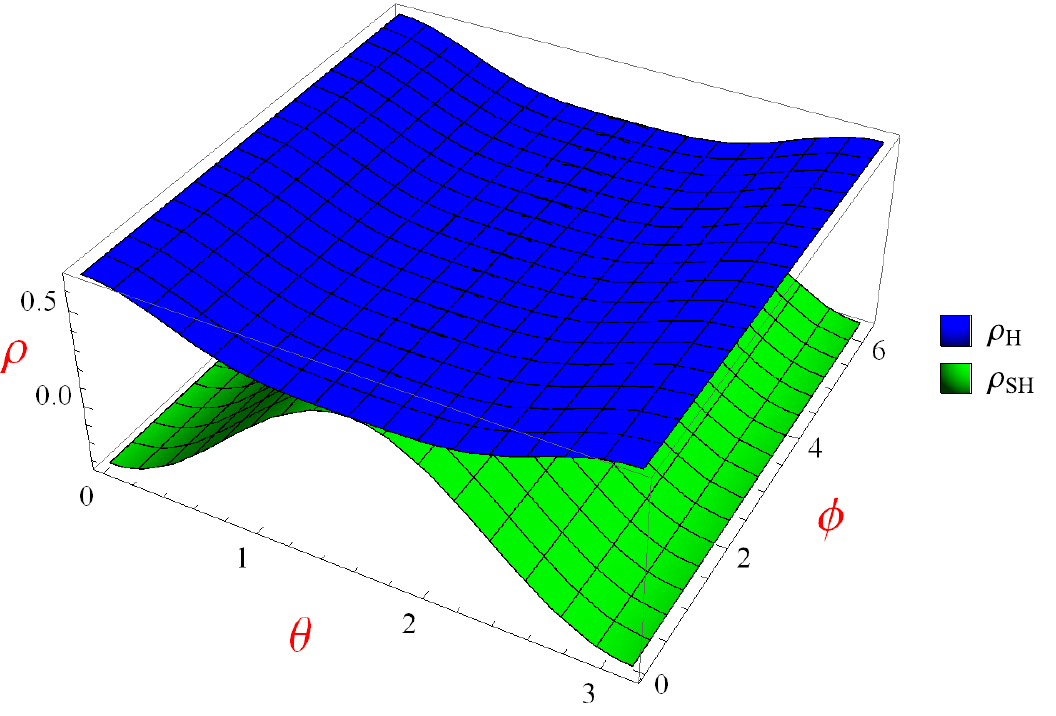}
\end{minipage}\hfill
\begin{minipage}{0.45\textwidth}
\centering
\includegraphics[scale=0.70]{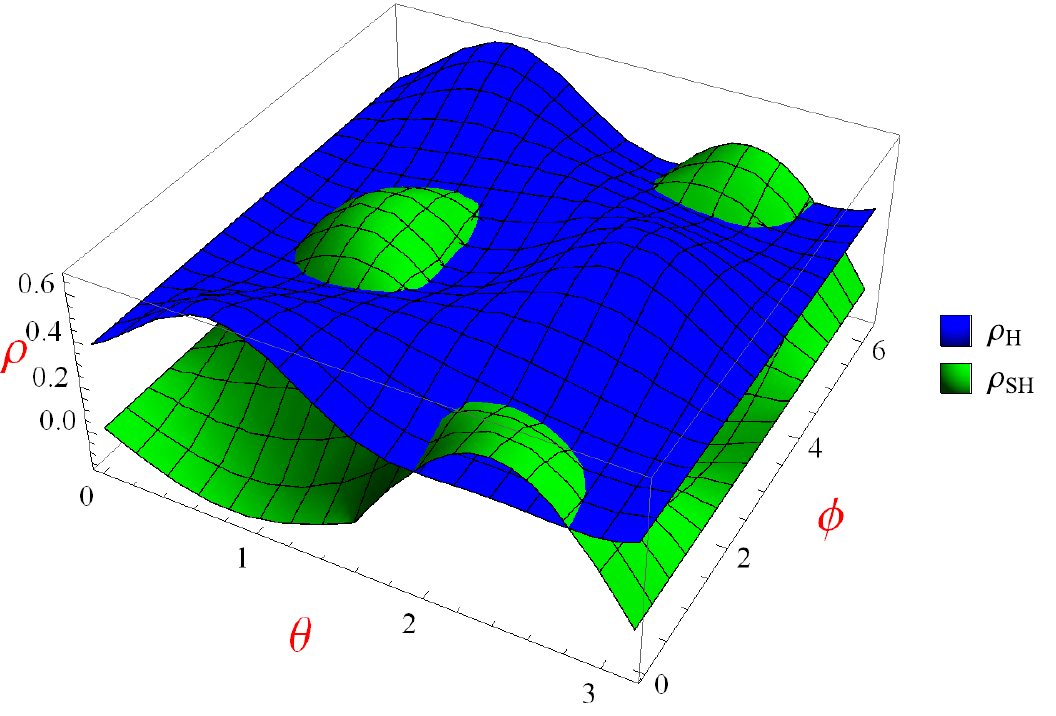}
\end{minipage}
\caption{Relative positions of the supertranslation horizon and the Killing horizon. On the left figure: $C(\theta,\phi)=\frac{M}{12}(3\cos^2\theta-1) $ ($\alpha = \frac{1}{2}$ in \eqref{toy1} ). The supertranslation horizon is entirely shielded by the Killing horizon. On the right figure:  $C(\theta,\phi)=\frac{M}{6} \sin 2\theta \cos\phi $. This corresponds to the toy model \eqref{toy2} with $\alpha=1$.  Depending on the angle, one first encounters the Killing horizon or the supertranslation horizon.
}\label{fig2}
\end{figure}

Assuming the bound \eqref{bound}, there are only two scenarios: starting from a specific angle $(\theta,\phi)$ and decreasing the radius $\rho$, either one first reaches the infinite redshift surface/Killing horizon $\rho_{H}(\theta,\phi)$ or either one first reaches the supertranslation horizon $\rho_{SH}(\theta,\phi)$. These horizons delimit the range of coordinates spanned by the metric \eqref{cgm}. Since there are two roots in \eqref{rhoHeq}, there are two branches for $\rho_H$. The highest branch is the relevant one since it is the first coordinate singularity that is hit starting from the asymptotic region. We illustrate in Figures \ref{fig1} and \ref{fig2} these two distinct scenarios.

\subsection{Conserved charges}

The spacetime is uniquely described by the total mass $M$ and the supertranslation field $C(\theta,\phi)$. Time translation invariance implies that the zero mode $C_{(0,0)}$ is irrelevant. Let us now characterize the spacetime instead by its conserved canonical charges. It was proven that in the case $M=0$ supertranslations are symplectic symmetries \cite{Compere:2016jwb}: these are asymptotic symmetries at null infinity which extend in the bulk and whose conserved charges can be evaluated on any surface in the bulk. We expect that the proof can be generalized for $M$ arbitrary since the addition of the mass does not introduce non-conserved or non-integrable charges. Asymptotic supertranslation charges can be computed in the null asymptotic region and lead to the charge
\bea
Q_{\chi_T}^S=\frac{M}{G}\oint_S d^2\Omega \, T(\theta,\phi)
\eea
where $\chi_T = T(\theta,\phi)\p_u+\dots$ is the generator of supertranslations. These charges all vanish except the zero mode which gives the total mass $M/G$, consistently with \cite{Flanagan:2015pxa,Hawking:2016msc}. Supertranslation symmetries except time translations are therefore trivial. The property that the Killing charge is 
constant on the orbit generated by the action of symplectic symmetries was also noted in three dimensional toy models \cite{Compere:2015knw}. 

Other symplectic symmetries exist in the case $M=0$: superrotation charges which were first computed by Barnich and Troessaert for general asymptotically flat spacetimes \cite{Barnich:2011mi} from the canonical Barnich-Brandt charge \cite{Barnich:2001jy} and which were further analysed in \cite{Compere:2016jwb}. In the case where the news vanishes as here, the charge is integrable and conserved and after discarding boundary terms (which defines a prescription) one gets
\bea
Q_{\chi_R}^S= \frac{1}{4G}\oint_S d^2\Omega \, R^A \left( 2 N_A + \frac{1}{16} D_A (C^{BC}C_{BC}) \right). 
\eea
The angular momentum aspect $N_A$ can be obtained by rewriting the metric in asymptotic BMS gauge at future null infinity. It reads as
\bea
N_A = - \frac{3}{32}D_A (C_{BC}C^{BC}) - \frac{1}{4} C_{AB} D_C C^{BC} + 3 M D_A C. 
\eea
The first two terms were obtained in the vacuum case \cite{Compere:2016jwb}. There is only one new term linear in the mass $M$. In the case $M=0$, the rotation and boost charges vanish \cite{Compere:2016jwb}. In this case where $M \neq 0$, the rotation charges still vanish because rotations are associated with $R^A$ which are Killing vectors and therefore obey $D_A R^A = 0$. Rotations are Killing symmetries of the metric \eqref{cgm} whose components take an unusual form because of the coordinate transformation \eqref{diffeo}. Again, applying a finite diffeomorphism whose infinitesimal generator is a symplectic symmetry does not affect the charges of Killing symmetries, here the $SO(3)$ rotation charges which are still identically zero.

The boost charges are generically non-vanishing. They depend upon the lowest $l=1$ harmonics of the supertranslation field $C$ and could therefore be attributed to a change of the center-of-mass as in any standard classical system. We conveniently choose the origin of the coordinate system to equal to position of the center-of-mass of the system by requiring that the 3 boost charges vanish. This sets the $l=1$ first harmonics of $C$ to zero. 

The non-trivial superrotation charges encode information about the supertranslation field. The supertranslation field appears quadratically and therefore its sign is not fixed by the conserved charges. It is not  clear whether the supertranslation conserved charges are sufficient to fully reconstruct $C$ up to a global sign.\footnote{An alternative approach was proposed in \cite{Campiglia:2015yka} where charges associated with diffeomorphisms on the sphere were considered. However, we obtain that the Barnich-Brandt charge \cite{Barnich:2001jy} then linearly diverges in $r$ at large radius as $\propto r \int_S d^2\Omega C (D^2+ 2) D_A R^A $. This divergence vanishes for superrotations but not for a general diffeomorphism on the sphere $\xi = R^A(\theta,\phi) \p_A + \dots$. We are therefore led to discard such charges.}  The existence of generically non-trivial superrotation charges confirms that the metric \eqref{cgm} is physically distinct from the Schwarzschild metric.

Note that in the static gauge, it is natural to define the gauge-dependent local charge
\bea
Q_L[f] = \frac{1}{8 \pi G} \oint_{SH} f dA\label{localQ}
\eea
where $SH$ is the supertranslation horizon, $dA = (1+\frac{M}{2\rho_s(\rho_{SH})})^2\sqrt{\text{Det}(\gamma^{SH}) }d \theta d\phi$ is the measure on $SH$ and $f(\theta,\phi)$ is an arbitrary function which is integrable with respect to the measure $dA$. In 3 spacetime dimensions, one can rewrite this local charge as a canonical charge associated with superrotation symmetries, as shown in Appendix \ref{threed}. In four dimensions, it is not clear whether this local charge could be rewritten as a symplectic symmetry.  

\subsection{Experiments on a finite patch or on the complete sphere}

The final question that we will address is whether or not the metric \eqref{cgm} leads to deviations to standard results in general relativity such as the bending of light. The metric is obtained by a regular diffeomorphism outside of the supertranslation horizon surface. One could apply any gauge transformation (coordinate transformation) which preserves the canonical charges to describe the metric. For example, one can always choose a coordinate patch in a finite solid angular range where the metric will be given by the Schwarzschild metric. 
Now,  one cannot find coordinates where the metric is Schwarzschild on the complete spherical range since superrotation charges which enclose the supertranslation horizon are non-trivial. 
The largest coordinate patch where one could write the metric as the Schwarzschild metric is necessarily bounded by a solid angle around the central object. In these special coordinates, the solid angle left with the supertranslation field would contain all information about the non-trivial conserved superrotation charges. There is however nothing special about that solid angle since one could find coordinates where the metric is Schwarzschild around any solid angle.\footnote{This reasoning bears some ressemblance with the unobservability of Dirac strings in the case of a global magnetic monopole \cite{Dirac:1931kp}.}  Therefore all experiments  in a finite solid angular range outside of the supertranslation horizon are unaffected by the supertranslation field. Detecting the supertranslation field requires an experiment which considers the entire sphere around the central object, for example by placing an array of rulers around the central object and deducing the integrated superrotation charge.

\section{Conclusions and Outlook}

We propose that the generic static final state of collapsing matter and radiation is not described by the Schwarzschild metric but instead by the metric \eqref{cgm} which possesses an asymptotic supertranslation field characterized by its superrotation charges. For small supertranslation field, the metric describes a black hole which finitely deviates from the Schwarzschild black hole. Large supertranslation fields are prohibited by the cosmic censorship hypothesis. Up to a caveat, the value of the supertranslation field can be obtained from the angle-dependent energy balance of the collapse and subsequent evolution of the black hole. Toy models were analysed but a detailed astrophysical model for neighbouring black holes Sgr A* and M87 would be required to accurately predict their metric. We argued that the deviation from the Schwarzschild metric does not lead to observable effects for experiments outside the so-called supertranslation horizon except for affecting global properties on the entire sphere around the black hole.

The boundary conditions that we used for deriving the angle-dependent conservation of energy are crucial for our arguments and would need to be assessed by independent considerations. The bound on the supertranslation field that we derived from the cosmic censorship hypothesis is testable from numerical simulations of black hole collapse. In simplified models, its proof or falsification could be obtained by deriving analytic boundedness properties of a specific differential operator of the sphere. 

The existence of these classical states weakens the black hole information paradox since classical black holes are characterized by additional conserved charges. Yet, much of the classical and quantum structure of the final states with supertranslation memories remains to be understood.

\section*{Acknowledgments}

G.C. would like to thank A. Strominger and his entire team, especially T. He, P. Mitra and S. Pasterski, for sharing their views and enthusiasm on BMS symmetry and memory effects during his visit at the CMSA center at Harvard in April 2015 as well as S. Doeleman, S. Gralla, M. Johnson, A. Lupsasca, M. Rodriguez and A. Strominger for an inspiring meeting on the Event Horizon Telescope. G.C. would also like to thank M. Guica for stimulating discussions. G.C. and J.L. both acknowledge the current support of the ERC Starting Grant 335146 HoloBHC ``Holography for realistic black holes and cosmology''. 
G.C. is Research Associate of the Fonds de la Recherche Scientifique F.R.S.-FNRS (Belgium). This work is also partially supported by FNRS-Belgium (convention IISN 4.4503.15).
\appendix

\section{BMS finite diffeomorphisms}
\label{Diff1}

This appendix is aimed at presenting one derivation of the diffeomorphism \eqref{diffeo} which generates a supertranslation field  profile labelled by an arbitrary smooth function $C(z,\bar z)$ on the sphere. There might be different ways to arrive at the final result. Here we present the path that we originally followed. In a crucial technical step, as we will see, we will use the existence of the enhanced BMS algebra with Weyl transformations \cite{Barnich:2009se,Barnich:2010eb,Barnich:2011mi} (see also the recent work \cite{Barnich:2016lyg}). The enhanced BMS algebra also contains singular-valued superrotations and singular-valued supertranslations \cite{Barnich:2009se,Barnich:2010eb,Barnich:2011mi,Barnich:2016lyg} which readily carry through the computation once one extends the globally Killing vectors on the round sphere which define the $SO(3,1)$ algebra to the locally defined Killing vectors which define the two-dimensional conformal algebra \cite{Barnich:2009se,Barnich:2010eb}. This appendix also allows to derive from first principles the vacuum metric with finite superrotation field dubbed ``vacua with sources'' presented in \cite{Compere:2016jwb}.

Let us start with the global Minkowski vacuum written in retarded and stereographic coordinates,
\be
ds^2=-du_s^2-2du_sdr_s+\frac{4r_s^2}{(1+z_s\bar{z}_s)^2}dz_sd\bar{z}_s\label{s2}. 
\ee
The metric falls into the gauge considered by Bondi, van der Burg, Metzner and Sachs \cite{Bondi:1962px,Sachs:1962wk}. The canonical infinitesimal diffeomorphism which generates BMS supertranslations and Lorentz transformations reads as \cite{Bondi:1962px,Sachs:1962wk} 
\bea
\chi_{T,R}^{(BMS_+)} &=&(T+\frac{1}{2}uD_AR^A)\partial_u-\frac{1}{2}r( D_AR^A-\frac{1}{r}D_AD^A(T+\frac{1}{2}uD_CR^C)+\mathcal{O}(\frac{1}{r^2}))\partial_r+\nonumber\\&&(R^A-\frac{1}{r} {D}^A(T+\frac{1}{2}uD_BR^B)+\mathcal{O}(\frac{1}{r^2}))\partial_A. \label{ST}
\eea
Here $T = T(z,\bar z)$ is an arbitrary real smooth function on the sphere and $R^A=(R(z),\bar{R}(\bar{z}))$ are the globally defined conformal Killing vectors on the unit round sphere.  The subleading terms in the radial expansion are uniquely fixed in BMS gauge. Around the Minkowski vacuum this radial expansion exactly stops at the order indicated but it is infinite around Minkowski spacetime transformed by an arbitrary finite diffeomorphism which preserves the BMS gauge. 

The problem at hand is to find the corresponding finite diffeomorphism at each order in the radial expansion and then resum it to obtain a closed form expression. The task is not straightforward since an arbitrary function $T(z,\bar z)$ is present, the metric and connection on the sphere matters as well and the radial expansion is only known so far as a series expansion.

In order to facilitate this daunting task, it is in fact very useful to first consider a new system of coordinates. Minkowski spacetime can be foliated by complex plane sections as
\bea
ds^2  = -2 du_c dr_c +  2 r_c^2 dz_c d\bar z_c. \label{cm}
\eea
The two metrics \eqref{s2} and \eqref{cm} are related by the finite coordinate transformation
\bea
r_c&=& \frac{\sqrt{2}r_s}{1+z_s\bar{z}_s}+\frac{u_s}{\sqrt{2}},\nn\\
u_c&=&\frac{1+z_s\bar{z}_s}{\sqrt{2}}u_s-\frac{z_s\bar{z}_su_s^2}{2 r_c},\label{ctos}\\
z_c&=&z_s-\frac{z_su_s}{\sqrt{2}r_c},\qquad  \bar{z}_c=\bar{z}_s-\frac{\bar{z}_su_s}{\sqrt{2}r_c}.\nn
\eea
One can uniquely invert this coordinate transformation as
\bea
r_s&=& \frac{1}{\sqrt{2}} \sqrt{ (u_c + r_c (1+z_c \bar z_c))^2 - 4 r_c u_c },\nn\\
u_s&=&  \frac{1}{\sqrt{2}}(u_c + r_c(1+ z_c \bar z_c)) - r_s,\label{stoc}\\
z_s&=& \frac{\frac{1}{\sqrt{2}}(u_c - r_c(1+ z_c \bar z_c) ) + r_s}{\bar z_c u_s },\qquad
\bar{z}_s=  \frac{\frac{1}{\sqrt{2}}(u_c - r_c(1+ z_c \bar z_c) ) + r_s}{z_c u_s } . \nn
\eea

The infinitesimal diffeomorphism generator of BMS and Weyl transformations has been written as an asymptotic series expansion in \cite{Barnich:2009se,Barnich:2010eb,Barnich:2011mi,Barnich:2016lyg}. It takes a simple form around the vacuum metric \eqref{cm} because there is no angular dependence. The task is then reduced to find the finite combined BMS transformation and Weyl rescaling around the metric \eqref{cm}. After some algebra involving a guess for the finite transformation, we were able to find the right coordinate change. It reads as 
\bea
r_c&=& \frac{\p_z \p_{\bar z} W}{\p_z G\bar{\p}G} +\sqrt{\frac{r^2}{(\p_u W)^2} +\frac{(\p_z^2G\p_zW-\p_z G\p_z^2 W)(\p_{\bar{z}}^2\bar{G}_0\p_{\bar{z}}W-\p_{\bar{z}}\bar{G}_0 \p_{\bar z}^2 W)}{(\p_zG)^3(\p_{\bar{z}}\bar{G})^3}},\nn\\
u_c&=& W - \frac{\p_z W \p_{\bar z} W }{\p_zG\p_{\bar{z}}\bar{G}r_c},\label{ctoBMS}\\
z_c&=&G- \frac{\p_{\bar z} W}{\p_{\bar{z}}\bar{G}r_c},\qquad  \bar{z}_c = \bar{G}- \frac{\p_{z} W}{\p_z Gr_c},\nn
\eea
where $W(u,z,\bar z)$ is an arbitrary function of $u,z,\bar{z}$, $G(z)$ is a combination of $1,z,z^2$ and $\bar G(\bar z)$ is its complex conjugate. The function $W$ characterizes the Weyl rescalings and supertranslations. The function $G(z)$ characterizes the Lorentz transformations. One can generalize $G(z)$ to be an arbitrary meromorphic function, which then generates superrotations. One can also generalize $W(u,z,\bar z)$ to be a singular function on the sphere which then generates singular Weyl rescalings and singular supertranslations. 

The resulting metric in $(u,r,z,\bar z)$ coordinates takes the asymptotic BMS form as considered by Sachs \cite{Sachs:1962zza} and generalized by Penrose \cite{Penrose:1962ij}. It reads as 
\bea
ds^2 = e^{2 \beta} \frac{V}{r}du^2 - 2 e^{2\beta} du dr + g_{AB}^{(BMS)}(dz^A - U^A du)(dz^B - U^B du)
\eea
where the asymptotic fall-off conditions of all fields are clearly detailed by Barnich and Troessaert \cite{Barnich:2010eb}. In particular the metric $g_{AB}^{(BMS_+)}$ takes the form 
\bea
g_{AB}^{(BMS_+)} = r^2 e^{2 \varphi} \gamma^c_{AB} + O(r)
\eea
where $\gamma^c_{AB}dz^A dz^B  = 2 dz d\bar z$ is the flat metric on the complex plane of the seed solution \eqref{cm} and the new conformal factor $\varphi$ depends upon $W$ and $G$ and is therefore time-dependent. Here, we are interested in recovering the time-independent conformal factor $e^{2 \varphi(z,\bar{z})} = \gamma_{z\bar z} = \frac{2}{(1+z \bar z)^2}$ fixed by Dirichlet boundary conditions. This forces to choose 
\bea
W =e^{-\varphi(z,\bar{z})}(u+C(z,\bar{z}))\sqrt{\p_zG\p_{\bar{z}}\bar{G}}, \label{WW}
\eea
One can therefore generate the vacuum with an arbitrary BMS supertranslation field profile $C(z,\bar z)$ and superrotation profile $G(z)$ in $(u,r,z,\bar z)$ coordinates by starting with the global Minkowski spacetime \eqref{s2} written in $(u_s,r_s,z_s,\bar z_s )$ coordinates, then using first the transformation \eqref{stoc} to switch to $ (u_c,r_c,z_c,\bar z_c )$ coordinates then \eqref{ctoBMS} to reach the $ (u,r,z,\bar z)$ coordinates. This exactly results in the metric (19) of \cite{Compere:2016jwb}.

The transformation \eqref{ctoBMS} with $W$ given by \eqref{WW}, $G(z) = z$, $C = 0$ and $e^{2\varphi} = \gamma_{z\bar z}$ exactly reduce to the transformation law \eqref{ctos} after renaming the final coordinates as $(u_s,r_s,z_s,\bar z_s)$. This is how \eqref{ctos} was originally derived.

\subsection{Supertranslation diffeomorphism in static coordinates}
\label{app:sc}

Let us specialize to supertranslations only by setting the superrotation field to the identity $G(z) = z$. We then have 
\bea
W = \frac{1+ z \bar z}{\sqrt{2}}(u + C(z, \bar z))
\eea
and we obtain the vacuum metric discussed in \cite{Compere:2016jwb}
\bea\label{vac1}
ds^2&=&-du^2-2dud(\sqrt{r^2+U}  + \frac{1}{2}(D^2+2)C)+ \left( (r^2 + 2U)\gamma_{AB} + \sqrt{r^2+U}  C_{AB} \right)dz^Adz^B\nn
\eea
where $C_{AB}=-(2D_AD_B-\gamma_{AB} D^2)C$ and $U=\frac{1}{8}C_{AB}C^{AB}$. One can check that under a supertranslation \eqref{ST} generated by $T(z,\bar z)$, $R^A = 0$, the metric remains in BMS gauge and changes according to the shift $\delta_T C(z,\bar z) = T(z,\bar z)$. 

Static coordinates are obtained by setting
\bea
t &=& u + \rho,\\
\rho &=& \sqrt{r^2+U} +E,\qquad E(z,\bar z) \equiv \frac{1}{2}D^2C +C - C_{(0,0)}.
\eea
As explained in \cite{Compere:2016jwb}, the shift of $\rho$ by the lowest (constant) spherical harmonic $C_{(0,0)}$ is fixed by requiring the invariance of the radius under a constant time shift $\delta C = \Delta t$. The metric then reads
\bea\label{met5}
ds^2&=&-dt^2 +d\rho^2 + \left( ((\rho- E)^2 + U)\gamma_{AB} + (\rho - E) C_{AB} \right)dz^Adz^B.
\eea
This metric can be compared with the original global Minkowski vacuum written in static coordinates
\be\label{met6}
ds^2=-dt_s^2 +d\rho_s^2 + \rho_s^2 \gamma_{AB} dz_s^A dz_s^B
\ee
where $t_s = u_s + \rho_s$.  Following the chain of coordinate transformations, we can finally relate these two coordinate systems by the change of coordinates
\bea
t_s &=& t + C_{(0,0)}, \nn \\
\rho_s &=& \sqrt{(\rho - C + C_{(0,0)})^2 + D_A C D^A C}, \label{diffeo2}\\
z_s &=& \frac{(z- \bar z^{-1}) (\rho - C + C_{(0,0)}) + (z + \bar z^{-1}) (\rho_s - z \p_z C - \bar z \p_{\bar z} C)}{2(\rho - C +C_{(0,0)}) + (1+z \bar z)(\bar z \p_{\bar z} C - \bar z^{-1} \p_z C)}.\nn
\eea
In that sense,  \eqref{diffeo2} is the supertranslation generating coordinate transformation. The equality between \eqref{met5} and \eqref{met6} under \eqref{diffeo2} is identical to the equality \eqref{trands3} using \eqref{diffeo}, which proves the statement in the main text.

The metric \eqref{met5} is written in static gauge defined as $g_{\rho A} = 0$, $g_{\rho\rho} = 1$. The generator of supertranslations in that gauge can be written as 
\be
\xi^{(stat)}_T=T_{(0,0)}\partial_t-(T-T_{(0,0)})\partial_{\rho}+\frac{C^{AB}D_BT-2D^AT(\rho-\frac{1}{2}(D^2+2)(C-C_{0,0}))}{2((\rho-\frac{1}{2}(D^2+2)(C-C_{0,0}))^2-U)}\partial_A. \label{xis}
\ee
These generators exactly commute under the adjusted bracket defined in \cite{Barnich:2010eb}
\be
[\xi_1,\xi_2]_{ad} \equiv [\xi_1,\xi_2]-\delta_{\xi_1}\xi_2+\delta_{\xi_2}\xi_1.
\ee
Here, the variation $\delta_{\xi_1}$ acts on the field $C$ and its $p$-th derivative, $p=1,2,\dots$ as a derivative operator contracted with the $p$-th derivative of $\delta_{T} C(\theta,\phi) = T(\theta,\phi)$. As a consequence of these vanishing commutation relations, the supertranslations act everywhere in the bulk spacetime described by the metric \eqref{met5} which extends the asymptotic result of  \cite{Barnich:2010eb}. In a group theory language, the metric \eqref{met5} describes the orbit of Minkowski spacetime under the supertranslation group. 

The relationship between the supertranslation symmetry generators in BMS gauge $\xi^{(BMS_+)}_T$ given in \eqref{ST} (with $R^A=0$) and the supertranslation symmetry generators in the static gauge $\xi^{(stat)}_T$ \eqref{xis} can be found to be
\bea
\xi^{(stat)}_T = \xi^{(BMS_+)}_T - \delta_{T} x_{(stat)}^\mu \frac{\p}{\p x^\mu_{(stat)}}.
\eea
where $x_{(stat)}^\mu = (t,\rho,\theta,\phi)$ are functions of $(u,r,\theta,\phi)$ which depend upon $C$. This allows to relate the BMS symmetry algebra between future null infinity and spatial infinity. 

\section{Supertranslations in 3-dimensional gravity}
\label{threed}

Let us consider the toy model of three dimensional Einstein gravity coupled to matter. Positive energy conditions on matter prevent the formation of an event horizon \cite{Ida:2000jh}. There is therefore no black hole in 3 dimensional asymptotically flat spacetimes. The best toy model that we could consider as a final state is a non-trivial supertranslation field alone. It is a solution to vacuum Einstein gravity to which we now turn.

The generic solution to vacuum Einstein gravity with zero cosmological constant with Dirichlet boundary conditions (i.e. the boundary metric is imposed to be the unit circle) and in a three-dimensional generalization of Bondi gauge was built in \cite{Barnich:2010eb}. This solution was understood as a limit of $AdS_3$ geometries in \cite{Barnich:2012aw}. It depends upon two arbitrary functions $\Theta(\phi)$ and $\Xi(\phi)$.  The field $\Theta$ transforms as the coadjoint representation under infinitesimal diffeomorphisms of the $\phi$ circle. It can be recognized as built from the superrotation field and a zero mode: the mass which cannot be generated by a diffeomorphism. If one keeps the mass arbitrary but discard the superotation field, one gets $\Theta = 8 G M$. Such a field leads to conical defects. Since we are not interested in conical defects for the following reasoning, we also set $M = - \frac{1}{8G}$.  The solution with the field $\Xi(\phi)$ alone reads as 
\bea
ds^2 = -du^2 - 2 du dr + 2 \Xi(\phi) du d\phi + r^2 d\phi^2.
\eea
The field $\Xi$ transforms under a supertranslation generated by $\chi_T = T(\phi)\p_u + \dots$ as $\delta_T \Xi =- T'''-T'$ where prime denotes a $\phi$ derivative and where we used the sign convention $\delta_T g_{\mu\nu} = +\mathcal L_{\chi_T}g_{\mu\nu}$ . It can therefore be recognized as a composite field in terms of the supertranslation field $C(\phi)$ and a zero mode, the angular momentum, as $\Xi = 4GJ-C'''- C'$. Here $C$ transforms as $\delta_T C = T$ and $J$ is the angular momentum which cannot be generated by a diffeomorphism. Since we are interested in the static case we set $J = 0$. Global Minkowski spacetime is recovered when $C = 0$. The metric reads as 
\bea
ds^2 = -du^2 - 2 du d( r +( C''+ C ) )  + r^2 d\phi^2.
\eea
Since the metric does not depend upon the lowest 3 harmonics of $C$, those do not contain any information. One can switch to static coordinates by defining $\rho = r + (C''+C-C_{(0)})$ and $t = u +\rho$ where $C_{(0)}$ is the zero mode in the harmonic decomposition of $C$. One gets
\bea
ds^2 = -dt^2 + d\rho^2+ (\rho - \rho_{SH}(\phi))^2 d\phi^2. \label{met3}
\eea
where we defined
\bea
\rho_{SH}(\phi) \equiv (\p_\phi^2 + 1)(C -C_{(0)}). 
\eea

The presence of the supertranslation field is signaled by the existence of superrotation charges associated with the infinitesimal diffeomorphisms on the circle $\xi = R(\phi)\p_\phi+ \dots $. The superrotation charge was computed in \cite{Barnich:2010eb} close to future null infinity $u$ fixed and $r \rightarrow \infty$ and reads as 
\bea
Q_R& =& \frac{1}{8 \pi G}\int_0^{2\pi} d\phi R(\phi) \Xi(\phi) =  \frac{1}{8 \pi G}\int_0^{2\pi} d\phi \p_\phi R(\phi) \rho_{SH}(\phi) \\
&=& \frac{1}{8 \pi G}\int_0^{2\pi} d\phi (C-C_{(0)}) \p_\phi (\p_\phi^2 + 1) R(\phi)\label{QR3}
\eea
after performing integration by parts. The superrotation charge vanishes when the supertranslation field reduces to a time translation $C_{(0)}$ or a spatial translation field generated by $C= a_x \cos\phi + a_y  \sin\phi$ with $a_x,a_y$ constants, consistently with the fact that these transformations are Killing symmetries. The last expression also shows that the 2 Lorentz charges ($R=\cos\phi$ and $\sin\phi$) and the rotation charge ($R=1$) are zero for a generic supertranslation field. There is a one-to-one correspondence between the $l \geq 2$ harmonics of the supertranslation field and the non-trivial superrotation charges.

Since the presymplectic structure of Einstein gravity vanishes \cite{Compere:2014cna}, superrotations are symplectic symmetries and the charges can be computed on any surface obtained by a smooth deformation. In particular, they can be computed around $\rho = \rho_{SH}$ which shows that there is a defect in spacetime located at either $\rho = \rho_{SH}$ or beyond that location.

In order to clarify the nature of the locus $\rho = \rho_{SH}(\phi)$, let us define $R = \rho-\rho_{SH}(\phi)$. The spatial metric reads as
\bea
ds^2 = (\rho'_{SH} d\phi+ dR)^2 + R^2 d\phi^2 = \left[ dR^2 + R^2 d\phi^2 \right] + 2 \rho'_{SH} d\phi dR + (\rho'_{SH})^2 d\phi^2. \label{md1}
\eea
The presence of the supertranslation field changes the metric by the two additional terms and the resulting metric is not the standard Euclidean metric. The induced metric on constant $R$ surfaces has the measure $\sqrt{g_{\phi\phi}} = |\rho_{SH}'|$ in the vicinity of $R=0$. This measure is generally non-zero and therefore the locus $R = 0$ is one-dimensional. Every periodic function has a derivative that vanishes at at least two points. At these points the measure of the induced metric on the locus $R=0$, namely $\sqrt{g_{\phi\phi}}$, vanishes. 

Let us study the metric in the coordinate region described by the solid angle $\phi = \phi_* + \delta \phi$ around an arbitrary angle $\phi_* $ where $\delta\phi$ is very small. The metric  \eqref{met3} can be expanded as 
\bea
ds^2 = d\rho_*^2 + \rho^2_* (d\delta \phi)^2 + O(\delta\phi^3). 
\eea
after defining the new radius $\rho_* = \rho - \rho_{SH}(\phi_*)$. For $\delta\phi$ small, the locus $\rho_* = 0$ is recognized as the origin of Euclidean space in polar coordinates where only a conical wedge is covered by $\delta \phi$. Therefore, one can define local Cartesian coordinates which extend the coordinate patch beyond $\rho_*=0$. The reasoning holds for all angles. In conclusion, the space does not end at the surface $R = 0$. It is a coordinate horizon where the static gauge $g_{\rho \phi} = 0$, $g_{\rho\rho} = 1$ has to break down.  We will call this circle the \emph{supertranslation horizon}. 

The structure of the supertranslation horizon can also be seen in Cartesian coordinates $X = R \cos\phi$, $Y = R \sin\phi$. This system of coordinates can be inverted in the region $\rho > \rho_{SH}$ in the standard way, $\phi = \arctan\frac{Y}{X}$, $\rho = \rho_{SH}(\phi) + \sqrt{X^2 + Y^2}$. After switching to Euclidean coordinates, we note that $g_{XX}g_{YY} - g_{XY}^2 = 1$ so the determinant of the metric no longer vanishes at the origin $X=Y=0$. The metric components read as
\bea
g_{XX} &=& \left( 1- \frac{XY}{(X^2+Y^2)^{3/2}} \rho_{SH}' \right)^2 + \frac{Y^4}{(X^2+Y^2)^3} (\rho_{SH}')^2,\\
g_{YY} &=&  \left( 1 + \frac{XY}{(X^2+Y^2)^{3/2}} \rho_{SH}' \right)^2 + \frac{X^4}{(X^2+Y^2)^3} (\rho_{SH}')^2. 
\eea
If one scales homogeneously $X \sim \eps$, $Y \sim \eps$, $dX \sim \eps$, $dY \sim \eps$, the spatial metric reduces to 
\bea
ds^2 = (\rho_{SH}')^2 \left( \frac{Y dX - X dY }{X^2 + Y^2}\right)^2 +O(\epsilon) =  (\rho_{SH}')^2 d\phi^2+O(\epsilon).\label{deg12}
\eea
The metric is well defined at all $X,Y$ real except $X=Y=0$. The Ricci scalar of the two-dimensional spatial metric is exactly zero outside of $X=Y=0$. The locus $X=Y=0$ is characterized by the profile function $\rho_{SH}(\phi)$. 

The data profile $\rho_{SH}(\phi)$ can also be obtained from a local charge defined at $R= 0$. Let us define a function $f(\phi)$ and a circle $C$ enclosing the locus $R=0$ at fixed small radius $R$. The local charge is defined as
\bea
Q_L[f] = \frac{1}{8 \pi G} \lim_{R \rightarrow 0} \oint_C f(s) ds. 
\eea
It can be evaluated as 
\bea
Q_L[f] = \frac{1}{8 \pi G} \int_0^{2\pi}  d\phi f(\phi) |\rho_{SH}'|. 
\eea
If one renames $f(\phi) = \text{sign}(\rho_{SH}') R(\phi)$, the local charge exactly coincides with the superrotation charge \eqref{QR3} which can be canonically associated with the asymptotic BMS supertranslation symmetry. 

The finite diffeomorphism which generates the metric \eqref{met3} can be found in closed form. The global Euclidean plane is given in terms of Cartesian coordinates $(x_s,y_s)$ and polar coordinates $(\rho_s,\phi_s)$ with $x_s = \rho_s \cos\phi_s$, $y_s = \rho_s \sin\phi_s$ as 
\bea
ds^2 = dx_s^2 + dy_s^2 = d\rho_s^2 + \rho_s^2 d\phi_s^2.\label{met4}
\eea
Equating the metrics \eqref{met3} with \eqref{met4} allows to find
\bea
x_s &=& (\rho - C(\phi)-C_{(0)}) \cos\phi + C'(\phi) \sin\phi, \\
y_s &=& (\rho - C(\phi)-C_{(0)}) \sin\phi -  C'(\phi) \cos\phi.  \label{chgt}
\eea
The Jacobian of the transformation $\text{Det}(\frac{\p x_s^i}{\p x^j})$ is $\rho - \rho_{SH}(\phi)$ and vanishes at the supertranslation horizon. Spatial translations generated by $C= a_x \cos\phi + a_y  \sin\phi$ correspond to shifting the coordinates as $x_s = x- a_x$, $y_s = y- a_y$ after defining the final Cartesian coordinates as $x=\rho\cos\phi$, $y=\rho \sin\phi$. The supertranslation horizon is a fixed locus under the action of Killing translations. Indeed, the Jacobian of the transformation \eqref{chgt} does not depend upon the $l=1$ harmonics of $C$ and is therefore translation invariant.  In polar coordinates, the diffeomorphism reads as 
\bea
\rho_s &=& \sqrt{(\rho - C +C_{(0)})^2 +(C')^2}\\
\phi_s &=& \arctan \left( \frac{(\rho - C +C_{(0)})\sin\phi - C' \cos\phi}{(\rho - C +C_{(0)})\cos\phi + C' \sin\phi}\right).
\eea
In Cartesian coordinates it reads as
\bea
x_s &=& x - C(\phi) \cos\phi + C'(\phi) \sin\phi, \\
y_s &=&y - C(\phi) \sin\phi -  C'(\phi) \cos\phi,  \label{chgt2}
\eea
where $\phi=\arctan\frac{y}{x}$.


\providecommand{\href}[2]{#2}\begingroup\raggedright\endgroup

\end{document}